\documentclass{emulateapj}
\usepackage{enumerate,pdfpages}
\begin{document}

\title{GOODS-{\em Herschel}: Impact of active galactic nuclei and star formation activity on infrared Spectral Energy Distributions at high redshift $^{\star}$}
\slugcomment{to be submitted to ApJ}

\author{Allison Kirkpatrick\altaffilmark{1}, Alexandra Pope\altaffilmark{1}, David M. Alexander\altaffilmark{2}, Vassilis Charmandaris\altaffilmark{3,4,5}, Emmanuele Daddi\altaffilmark{6}, Mark Dickinson\altaffilmark{7}, David Elbaz\altaffilmark{6}, Jared Gabor\altaffilmark{6}, Ho Seong Hwang\altaffilmark{8}, Rob Ivison\altaffilmark{9}, James Mullaney\altaffilmark{6}, Maurilio Pannella\altaffilmark{6}, Douglas Scott\altaffilmark{10}, Bruno Altieri\altaffilmark{11}, Herve Aussel\altaffilmark{6}, Fr\'ed\'eric Bournaud\altaffilmark{6}, Veronique Buat\altaffilmark{12},
Daniela Coia\altaffilmark{11},
Helmut Dannerbauer\altaffilmark{13}, Kalliopi Dasyra\altaffilmark{6}, Jeyhan Kartaltepe\altaffilmark{7}, Roger Leiton\altaffilmark{6,14}, Lihwai Lin\altaffilmark{15}, Georgios Magdis\altaffilmark{16}, Benjamin Magnelli\altaffilmark{17}, Glenn Morrison\altaffilmark{18,19}, Paola Popesso\altaffilmark{17}, Ivan Valtchanov\altaffilmark{11}}

\altaffiltext{$\star$}{{\it Herschel} is an ESA space observatory with science instruments provided by European-led Principal Investigator consortia and with important participation from NASA.}
\altaffiltext{1}{Department of Astronomy, University of Massachusetts, Amherst, MA 01002, USA, kirkpatr@astro.umass.edu}
\altaffiltext{2}{Department of Physics, Durham University, Durham DH1 3LE, UK}
\altaffiltext{3}{Department of Physics and Institute of Theoretical \& Computational Physics, University of Crete, GR-71003, Heraklion, Greece}
\altaffiltext{4}{IESL/Foundation for Research \& Technology-Hellas, GR-71110, Heraklion, Greece}
\altaffiltext{5}{Chercheur Associ\'e, Observatoire de Paris, F-75014,  Paris, France}
\altaffiltext{6}{Laboratoire AIM, CEA/DSM-CNRS-Universit{\'e} Paris Diderot, Irfu/SAp, Orme des Merisiers, F-91191 Gif-sur-Yvette, France}
\altaffiltext{7}{National Optical Astronomy Observatory, 950 North Cherry Avenue, Tucson, AZ 85719, USA}
\altaffiltext{8}{Smithsonian Astrophysical Observatory, 60 Garden Street, Cambridge, MA, 02138, USA}
\altaffiltext{9}{UK Astronomy Technology Centre, Royal Observatory, Blackford Hill, Edinburgh, EH9 3HJ, UK}
\altaffiltext{10}{Department of Physics \& Astronomy, University of British Columbia, 6224 Agricultural Road, Vancouver, BC V6T 1Z1, Canada}
\altaffiltext{11}{Herschel Science Centre, European Space Astronomy Centre, Villanueva de la Ca\~nada, 28691 Madrid, Spain}
\altaffiltext{12}{Laboratoire d'Astrophysique de Marseille -- LAM, Universit\'e d'Aix-Marseille, CNRS, UMR7326, 38 rue F. Joliot-Curie, 13388 Marseille Cedex 13, France}
\altaffiltext{13}{Universit\"at Wien, Institut f\"ur Astrophysik, T\"urkenschanzstra\ss e 17, 1180 Wien, Austria}
\altaffiltext{14}{Astronomy Department, Universidad de Concepci\'on, Casilla 160-C, Concepci\'on, Chile}
\altaffiltext{15}{Institute of Astronomy \& Astrophysics, Academia Sinica, Taipei 106, Taiwan}
\altaffiltext{16}{Department of Physics, University of Oxford, Keble Road, Oxford OX1 3RH, UK}
\altaffiltext{17}{Max-Planck-Institut f\"ur Extraterrestrische Physik (MPE), Postfach 1312, 85741, Garching, Germany}
\altaffiltext{18}{Institute for Astronomy, University of Hawaii, Manoa, HI 96822, USA}
\altaffiltext{19}{Canada–France–Hawaii Telescope Corp., Kamuela, HI 96743, USA}

\begin{abstract}
We explore the effects of active galactic nuclei (AGN) and star formation activity on the infrared (0.3 -- 1000$\,\mu$m) spectral energy distributions of luminous infrared galaxies from $z$ = 0.5 to 4.0.
We have compiled a large sample of 151 galaxies selected at 24$\,\mu$m ($S_{24}\gtrsim 100\,\mu$Jy) in the GOODS-N and ECDFS fields for which we have deep {\em Spitzer} IRS spectroscopy, allowing us to decompose the mid-IR spectrum into contributions from star formation and AGN activity.
A significant portion ($\sim$ 25\%) of our sample is dominated by an AGN ($>$ 50\% of mid-IR luminosity) in the mid-IR. Based on the mid-IR classification, we divide
our full sample into four sub-samples: $z\sim1$ star-forming (SF) sources; $z\sim2$ SF sources; AGN with clear 9.7$\,$\micron\ silicate absorption; and AGN with featureless mid-IR spectra.
From our large spectroscopic sample and wealth of multi-wavelength data, including deep {\em Herschel} imaging at 100, 160, 250, 350, and 500$\,$\micron,
we use 95 galaxies with complete spectral coverage to create a composite spectral energy distribution (SED) for each sub-sample. We then fit a two-temperature component modified blackbody to the SEDs. 
We find that the IR SEDs have similar cold dust temperatures, regardless of the mid-IR power source, but display a marked difference in the warmer dust temperatures. 
We calculate the average effective temperature of the dust in each sub-sample and find a significant
($\sim20\,$K) difference between the SF and AGN systems. We compare our composite SEDs to local templates and find that local templates do not accurately reproduce the mid-IR features and dust temperatures
of our high redshift systems. High redshift IR luminous galaxies contain significantly more cool dust than their local counterparts. 
We find that a full suite of photometry spanning the IR peak is necessary to accurately account for the dominant dust temperature components in high redshift IR luminous galaxies.
\end{abstract}

\keywords{Galaxies: evolution -- Galaxies: active -- Galaxies: star formation -- Infrared: galaxies -- ISM: dust}

\section{Introduction}
Understanding the link between star formation and active galactic nucleus (AGN) activity in the high redshift Universe is key to determining a complete picture of galaxy evolution.
The history of the star formation rate density in the Universe shows a peak in the range $z \sim$~1~--~3 \citep[e.g.,][]{bouwens2009}, during which time at least 50\% of the stars in the local Universe are expected to have been formed \citep[e.g.,][]{dickinson2003}. Furthermore, the bulk of the star formation during this peak period is occurring in luminous infrared galaxies (LIRGs, $L_{\rm{IR}}=10^{11}$--$10^{12}{\rm L}_{\odot}$) and ultra luminous infrared galaxies (ULIRGs, $L_{\rm{IR}}>10^{12}{\rm L}_{\odot}$) \citep[e.g.,][]{magnelli2011,murphy2011a}.
During the same epoch, the black holes within the centers of massive galaxies are building up their mass \citep{wall2005,kelly2010}, making this a key period for the growth of galaxies.
 
It is important to understand how each mechanism (star formation and AGN activity) manifests itself in observations of the interstellar medium, particularly the dust. 
In the far-infrared, dust grains emit thermal emission characterized by a blackbody spectrum with an additional $\nu^{\beta}$ term to account for the emissivity of the dust \citep[][]{hildebrand1983}. 
Each galaxy will have a distribution of dust temperatures depending on the size and distribution of the dust relative to the heating sources; the integrated IR spectral energy distribution (SED) will be a combination of modified blackbodies from all dust components. 
Most studies of high redshift galaxies fit a single modified blackbody template, assuming an average dust temperature, but there is evidence from local star-forming galaxies and ULIRGs that at least two dust components are needed to obtain a good fit to the full IR SED \citep{dunne2001,farrah2003,willmer2009,galametz11}. 
Physically, this can be understood through the main regions where dust can be found in a galaxy: cool diffuse (cirrus) dust in the interstellar medium; warmer dust in active star-forming regions; and even warmer dust found in an AGN torus \citep{rowanrobinson1989}. 

To study the properties of AGN and star-forming (SF) galaxies at high redshift, it is necessary to first identify systems likely harboring an AGN. Many previous studies have used X-ray detections to determine whether a
galaxy is hosting an AGN \citep[e.g.,][]{alexander2003}. However, as the majority of star formation at $z \sim$~1~--~3 occurs behind dust, many AGN will also be obscured by dust and gas and undetectable
in the X-ray \citep[e.g.,][]{alexander2008}. Therefore, we must rely on alternative methods which are insensitive to dust obscuration, to identify the presence and strength of an AGN. 
 
The mid-infrared spectrum, consisting of dust-reprocessed light, is rich in information about the underlying power sources of the galaxy. The mid-IR spectrum contains three major features that can be used as a
diagnostic of the mechanism driving the intense luminosity of the galaxy. First, the mid-IR spectrum exhibits polycyclic aromatic hydrocarbon (PAH) emission lines. PAHs are excited by UV and/or optical photons, most often produced by young stars; as such, PAHs are good tracers of the amount of star formation in a galaxy \citep[e.g.,][]{peeters2004,brandl2006}. Second, the mid-IR spectrum may exhibit additional continuum emission coming from hot dust around the AGN. \citet{mullaney2011} find that, on average, such emission can be represented as a steep power law ($\nu^{-2}$) below 20$\,\mu$m which flattens at longer wavelengths, falling off steeply beyond 40$\,\mu$m. The final main features observed in the mid-IR spectrum of galaxies are due to silicate absorption, most clearly seen at
9.7$\,\mu$m. This absorption feature can be due to either the AGN or the star formation, since it is produced by a dust screen surrounding a hot emission region, though there is some observational evidence that it is more prevalent in AGN with an optically thick dusty torus \citep{spoon2007}.

High redshift ($z=1-3$) (U)LIRGs are an attractive option for studying both SF and AGN activity in this peak epoch of stellar growth.
Not only are they undergoing intense periods of star formation (SFR~$\ge$~10 - 100~M$_\odot$ yr$^{-1}$),
but many show signs of concurrent AGN growth \citep[e.g.,][]{lutz2008,coppin2010}. With the advent of the space-based IR observatories {\em Spitzer Space Telescope} and {\em Herschel Space Observatory} \citep{pilbratt2010},
we are now in a position to obtain spectroscopy and photometry for these systems from near- to far-IR wavelengths, which will allow us to decompose the SEDs into the relative contributions
from AGN and SF activity. Quantifying each component in high redshift (U)LIRGs will enable us to investigate evolutionary trends of SF galaxies and AGN with redshift and determine how the two components might be affecting the observed dust properties

We have assembled a sample of 151 high redshift (U)LIRGs with deep {\em Spitzer} mid-IR spectroscopy. We use the mid-IR spectrum not only to determine the dominant mid-IR power source
(SF or AGN activity), but also to quantify the relative contribution of each to the mid-IR luminosity of every galaxy. We combine our spectroscopy with multi-wavelength data,
specifically new {\em Herschel} imaging at 100, 160, 250, 350, and 500$\,\mu$m, to study the
full IR SED of high redshift LIRGs and ULIRGs. We aim to study how mid-IR classification is related to far-IR properties such as dust temperature and IR luminosity. 

High redshift (U)LIRGs have long been taken to simply be analogs to local systems of the same luminosity. Due to the limited data previously available for high redshift (U)LIRGs, astronomers often resort to blindly applying local galaxy templates to high redshift systems to estimate parameters such as dust temperature and total IR luminosity. 
Recent studies have found that the SEDs of high redshift galaxies do not fit local templates, at least at $z>1.5$
\citep[e.g.,][]{pope2006,papovich2007,desai2007,dannerbauer2010,elbaz2010,nordon2010}. Given the considerable size of our sample with deep mid-IR spectroscopy and a wealth of
multi-wavelength data, we are in a unique position to create full IR SEDs for high redshift (U)LIRGs, test how accurately they reflect what we see in the local Universe, and determine where and when variations occur.

The paper is laid out as follows. In Section~2, we describe details of our sample and the mid-IR spectral decomposition of individual sources. In Section~3, we describe
how we created composite IR SEDs for four sub-samples. In Section~4, we discuss specific features of our new composite SEDs, and we compare these composites to local templates.
Finally, in Section~5 we discuss our results and summarize in Section~6.
Throughout this paper, we assume a standard cosmology with $H_{0}=70\,\rm{km}\,\rm{s}^{-1}\,\rm{Mpc}^{-1}$, $\Omega_{\rm{M}}=0.3$, and $\Omega_{\Lambda}=0.7$. 

\section{Data}

\subsection{Sample Selection}
Our sample consists of 151 high redshift galaxies from the Great Observatories Origins Deep Survey North (GOODS-N) and Extended Chandra Deep Field Survey (ECDFS) fields. We include all sources in these fields that were observed with the {\it Spitzer} IRS \citep[for details on the instrument, see][]{houck2004}.
While this sample contains a diverse range of sources, depending on the goals of each individual observing program, the overlying selection criterion is that each source is bright enough at 24$\,\mu$m to be observable with the IRS in a reasonable amount of time ($<$ 10 hours). As a result, 93\% of the sources have $S_{24}>100\,\mu$Jy, 80\% have $S_{24}>200\,\mu$Jy and 60\% have $S_{24}>300\,\mu$Jy. More details on this database of IRS sources in GOODS-N and ECDFS can be found in a separate paper (Pope et al.~in preparation).

\begin{figure*}
\plottwo{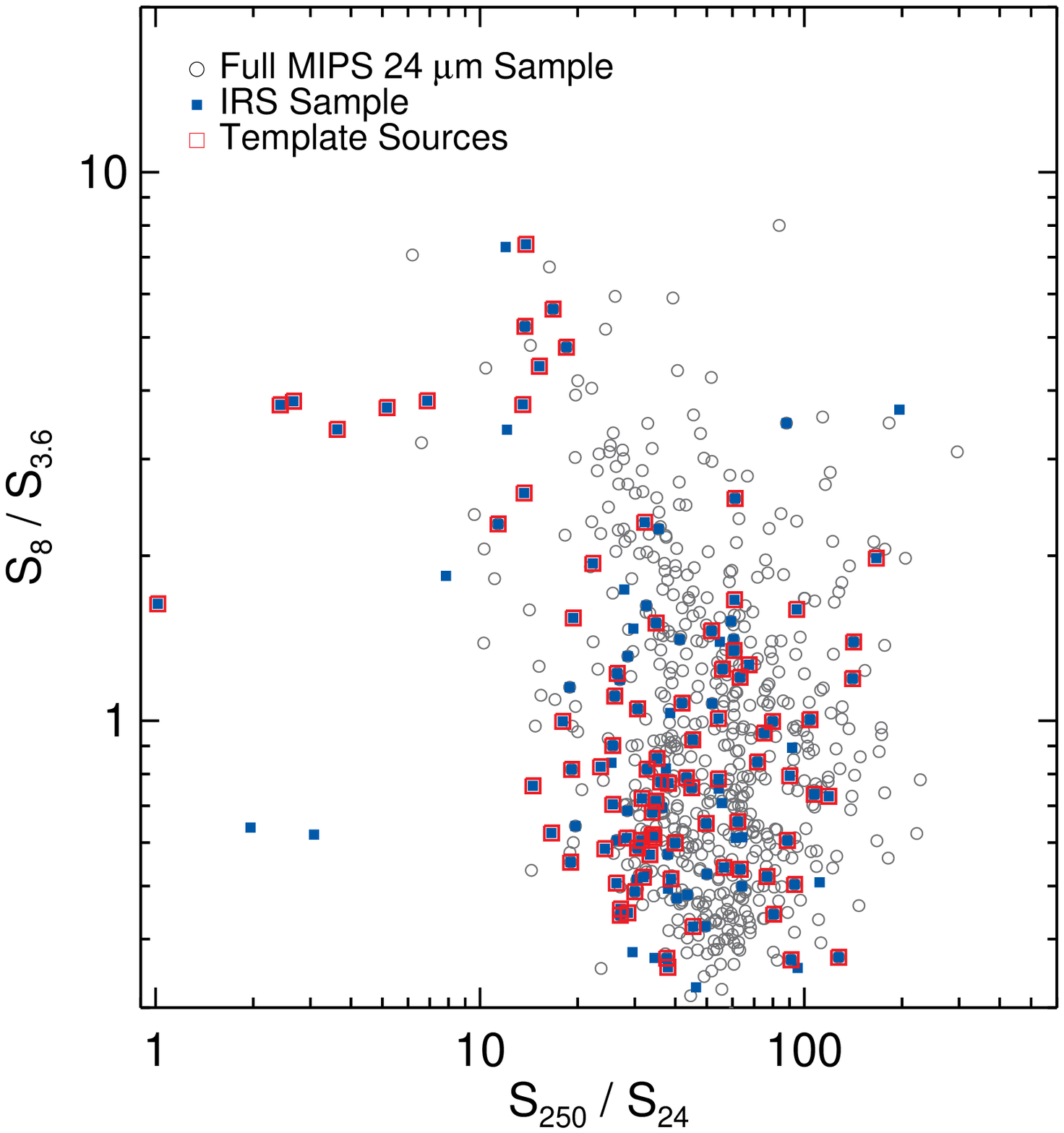}{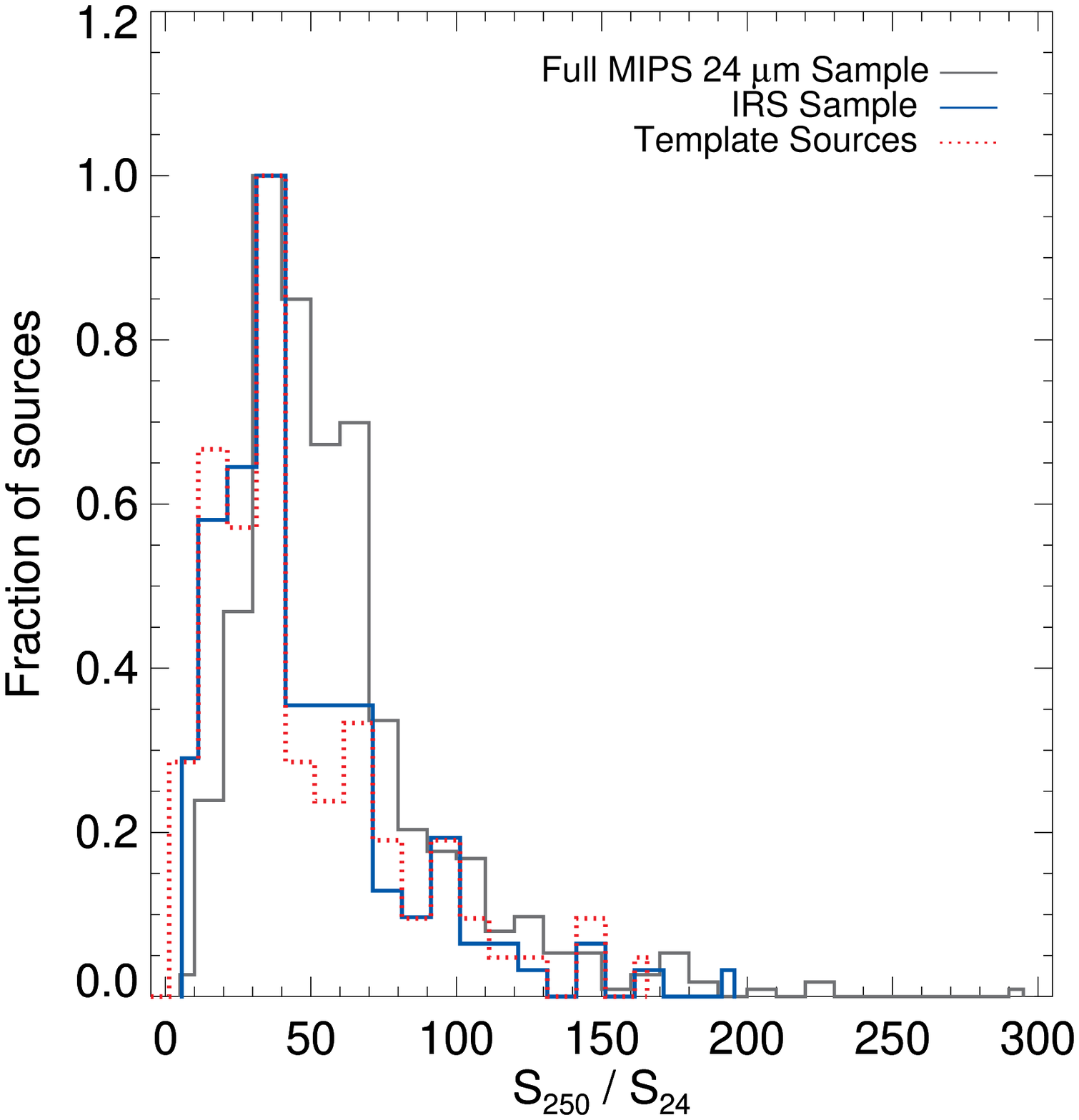}
\caption{The distribution of the MIPS 24\,$\mu$m GOODS sample with $S_{24}>100\,\mu$Jy  and $S_{250}>3\sigma$ and the IRS sample presented in this work in a colorspace combining far-, mid-, and near-IR fluxes (left) and
in the color $S_{250}/S_{24}$ (right). In \S \ref{sec:sed}, we use 95 sources from our IRS sample to create composite templates. We indicate the distribution of these 95 sources in red (see the online journal for a color version of this figure). The IRS sample follows the a similar distribution of colors as the full 24\,$\mu$m population with $S_{24}>100\,\mu$Jy.  \label{col_dist}}
\end{figure*}

It is important to determine how representative the IRS sample is of the full MIPS 24$\,\mu$m population.
Within GOODS-N, the IRS sources make up $\sim20\%$ of all 24$\,\mu$m sources above 200$\,\mu$Jy and $\sim30\%$ above 300$\,\mu$Jy.
If we limit ourselves to only 24$\,\mu$m sources with spectroscopic redshifts greater than 0.5, then the IRS sources make up $\sim40\%$ above 200$\,\mu$Jy and $\sim60\%$ above 300$\,\mu$Jy. 
In Figure \ref{col_dist}, we show that our IRS sample spans the same distribution in {\em Herschel+Spitzer} colors as the full MIPS population with $S_{24}>100\,\mu$Jy indicating that our IRS sample is representative within these parameters. In Section \ref{sec:sed}, we use 95 of our sources to create composite SEDs, and we show the distribution of these template sources in Fig. \ref{col_dist} as well. We run a Kolmogorov-Smirnov (KS) comparison test of the parent MIPS population and the two subpopulations (our full IRS sample and limited template sources). The significance level of the
KS test, for each of the subpopulations and the MIPS population, is 93\%; therefore, we cannot reject the null hypothesis that the populations are drawn from the same parent distribution.

As an additional test, we can determine if the fraction of AGN-dominated sources (defined using the mid-IR spectrum, see \S \ref{sec:decomp}) within our IRS sample is proportionate to that of all 24$\,\mu$m sources. 
In Kirkpatrick et al. (2012b, ApJ submitted), we develop new color-color diagnostics to separate AGN and SF sources at high redshift calibrated using this IRS sample. 
Using our new color cuts, we determine which 24\,$\mu$m sources from the full MIPS samples are AGN dominated in the mid-IR. The AGN dominated sources in our IRS sample have a flux density distribution consistent with the mid-IR color-selected AGN in the full MIPS sample above $S_{24}>100\,\mu$Jy. 
This gives us confidence that the IRS sample is providing fair and unbiased sampling of the full 24$\,\mu$m population above these flux limits. We report the {\em Spitzer} MIPS 24 and $70\,\mu$m photometry and {\em Herschel} PACS and SPIRE photometry for our full sample in Appendix A. 

Furthermore, it is important to determine the overlap between 24\,$\mu$m and 100\,$\mu$m selected galaxy samples to determine how representative our 24\,$\mu$m selected sample is of the greater population of IR luminous galaxies at high redshift. When looking at a blind catalog of sources selected at 100\,$\mu$m in GOODS-N with $S_{100}>1.1$\,mJy (the detection limit for this survey, see Elbaz et al.~2011), we find that 97\% of PACS 100\,$\mu$m are detected at 24\,$\mu$m \citep[see also][]{magdis2011}, and 67\% of PACS 100\,$\mu$m sources have $S_{24}>100\,\mu$Jy. Moreover, 70\% of our IRS sources have a detection at 100 $\mu$m.
Therefore, while we are selecting sources at 24\,$\mu$m in this study with $S_{24}>100\,\mu$Jy, we are getting a representative sampling of the bulk of the PACS 100\,$\mu$m selected sources. 
Given the large beam sizes at SPIRE wavelengths, which make robust counterpart identification challenging, we are unable to perform a similar simple test of the overlap of 24\,$\mu$m sources with those selected in SPIRE. However, we do note that the bulk of submm sources selected at even longer wavelengths, 850\,$\mu$m, are detected at 24\,$\mu$m and most of these detections are brighter than $S_{24}=100\,\mu$Jy \citep{pope2006}. 
It is important to keep in mind that in this paper, we focus on sources that have mid-IR spectra and PACS and/or SPIRE photometry. Therefore, our sample is representative of sources that are detected both in the mid-IR and far-IR and may not cover the parameter space of sources fainter than our flux limits or sources detected in either the mid-IR or the far-IR.

The low resolution ($R=\lambda/\Delta \lambda \sim 100$) {\it Spitzer} IRS spectra were reduced following the method outlined in \citet{pope2008a}. Specifically, since many of these are long integrations, we take care to remove latent build-up on the arrays over time, and we create a `supersky' from all the off-nod observations to remove the sky background. One dimensional spectra are extracted using the {\it Spitzer} IRS Custom Extraction (SPICE) in optimal extraction mode. For each target a sky spectrum is also extracted to represent the uncertainty in the final target spectrum.

\begin{deluxetable*}{lccccc}[h!]

\renewcommand{\tabcolsep}{0pt}
\tablecolumns{6}
\tablecaption{Basic properties of our four sub-samples. \label{basictbl}}
\tablehead{\colhead{} & \colhead{No. of} & \colhead{Median} & \colhead{Median $S_{8}$\tablenotemark{b}} & \colhead{Median $S_{24}$\tablenotemark{b}}
 & \colhead{Median $S_{100}$\tablenotemark{b}}\\
\colhead{Sub-sample} & \colhead{Sources\tablenotemark{a}} & \colhead{Redshift} & \colhead{($\mu$Jy)} & \colhead{($\mu$Jy)} &
 \colhead{(mJy)}}
\startdata
$z\sim1$ SF galaxies & 39 & 1.0 [0.8, 1.0] & \ 42   \ \  [23, 57] & \ 370 \ \ [260, 570] &  7.9  [4.8, 14.7]   \\
$z\sim2$ SF galaxies & 30 & 1.9 [1.8, 2.1] &  \ 17  \ \  [13, 38] & \ 270 \ \ [220, 370]  &  3.2  \ [1.6,\ 5.0]   \\
Silicate AGN 	   & 17 & 1.9 [1.6, 2.0] &  \ 64 \  [21, 140] & \ 470  \ \ [250, 860] &   5.3  [3.0, 10.2]  \\
Featureless AGN    & \ 9  & 1.2 [0.6, 1.6] &  288 [240, 311] & 1520 [1240, 2300]   &   9.5  [4.0, 11.7]

\enddata
\tablenotetext{}{We list the lower and upper quartile values in parenthesis next to each median value.}
\tablenotetext{a}{We list the number of sources in each sub-sample that are used to create the composite SEDs as well as the number of sources we have rejected in each subsample in Section \ref{sec:sed}.}
\tablenotetext{b}{Observed frame flux densities.}
\end{deluxetable*}

\subsection{Multi-wavelength Data}

The GOODS fields have been extensively surveyed and are rich in
deep multi-wavelength data, including: {\em Chandra} 2 Ms X-ray observations \citep{alexander2003, luo2008, xue2011}; 3.6, 4.5, 5.8, and 8.0$\,\mu$m imaging from the Infrared Array Camera (IRAC) on {\em Sptizer} (Dickinson 
et al.~in preparation);
IRS peak-up observations at 16$\,\mu$m \citep{teplitz2011}; and MIPS imaging at 24 and 70$\,\mu$m \citep{magnelli2011}. Recently, GOODS-N and GOODS-S have been surveyed with the GOODS-{\em Herschel} Open Time Key Program
\citep[P.I. David Elbaz,][]{elbaz2011} using both the PACS \citep{poglitsch2010} and SPIRE \citep{griffin2010} instruments, providing deep photometry at five far-IR and submm wavelengths: 100, 160, 250, 350, and 500$\,\mu$m. For the {\em Herschel} imaging, flux densities and the associated uncertainties were obtained by point source fitting using 24$\,\mu$m prior positions, allowing us to probe deeper limits
in the {\em Herschel} images. In addition, the GOODS-{\em Herschel} catalog is only comprised of sources with a clean detection, based on the 24$\,\mu$m prior position having
no bright neighbors in a given passband \citep[for further details, see][]{elbaz2011}.

We combine this space-based imaging with ground-based imaging in the near-IR ($J$ and $K$ bands) from VLT/ISAAC \citep{retzlaff2010} and CFHT/WIRCAM \citep{wang2010,lin2012}. At the longest (sub)mm wavelengths we use available data from LABOCA on APEX \citep{weiss2009} and the combined AzTEC+MAMBO mm map of GOODS-N \citep{penner2011}.

\subsection{Mid-IR Spectral Decomposition}
\label{sec:decomp}

\begin{figure*}
\begin{center}
\includegraphics[width=2.3in,angle=0]{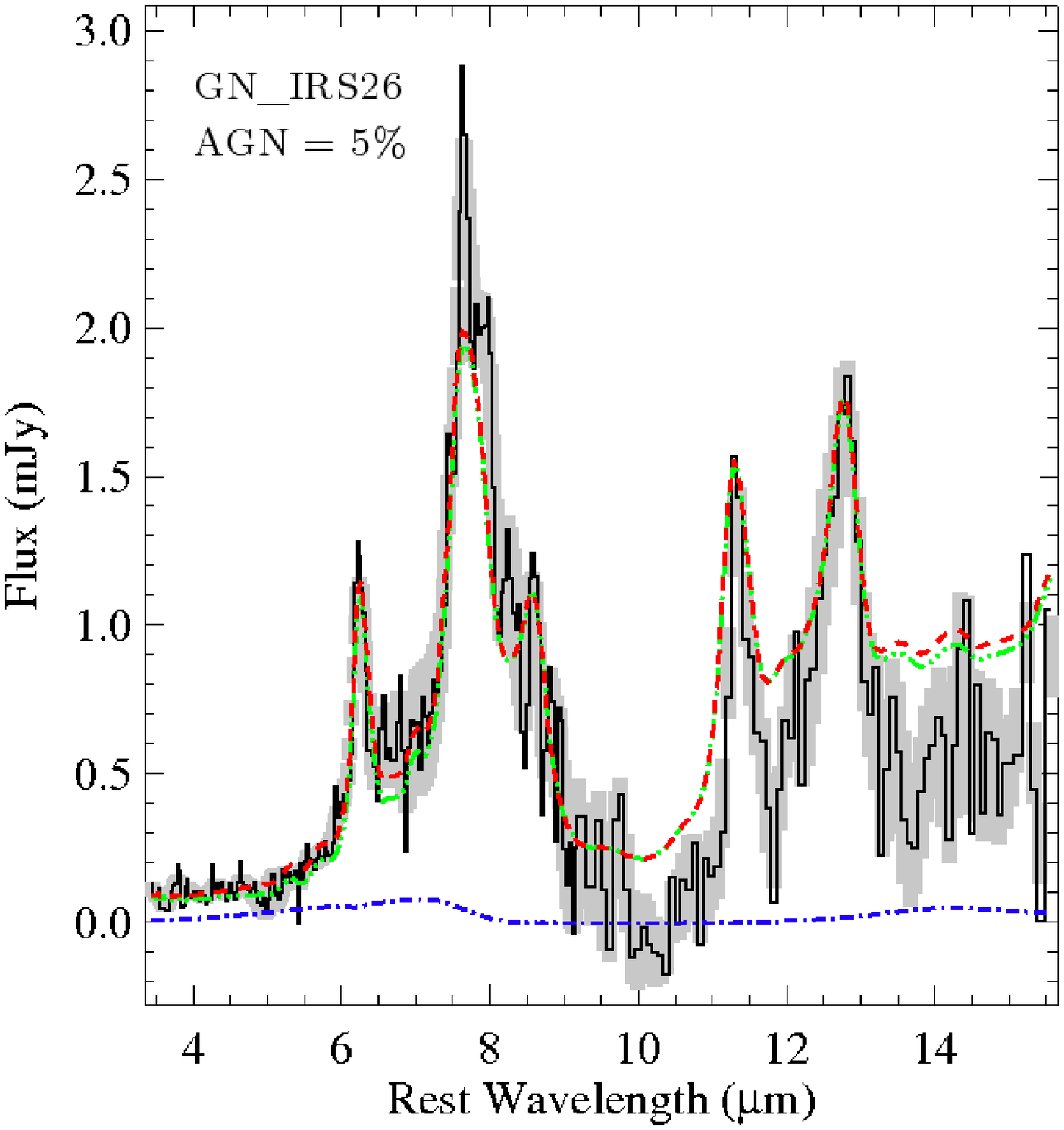}
\includegraphics[width=2.3in,angle=0]{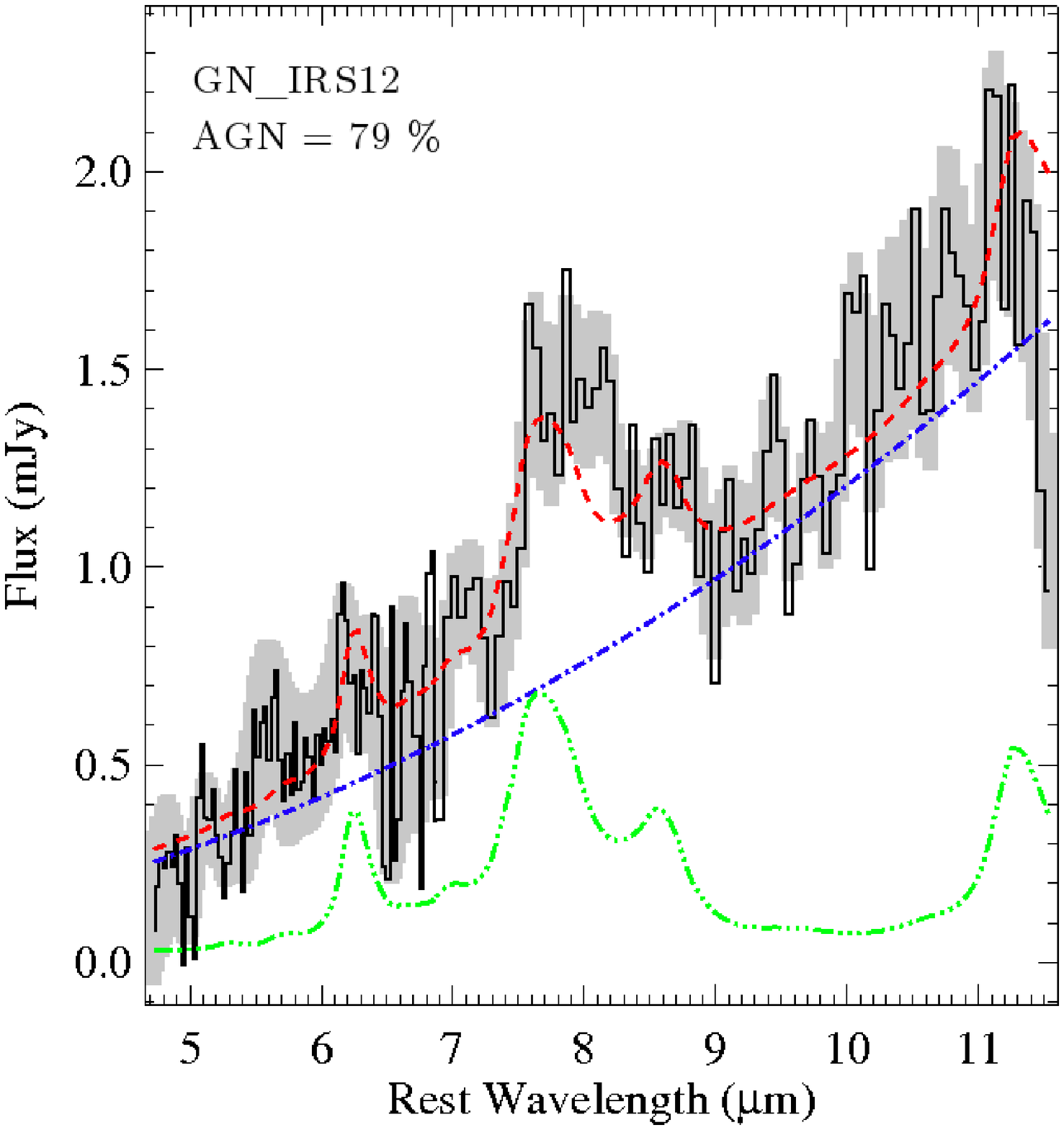}
\includegraphics[width=2.3in,angle=0]{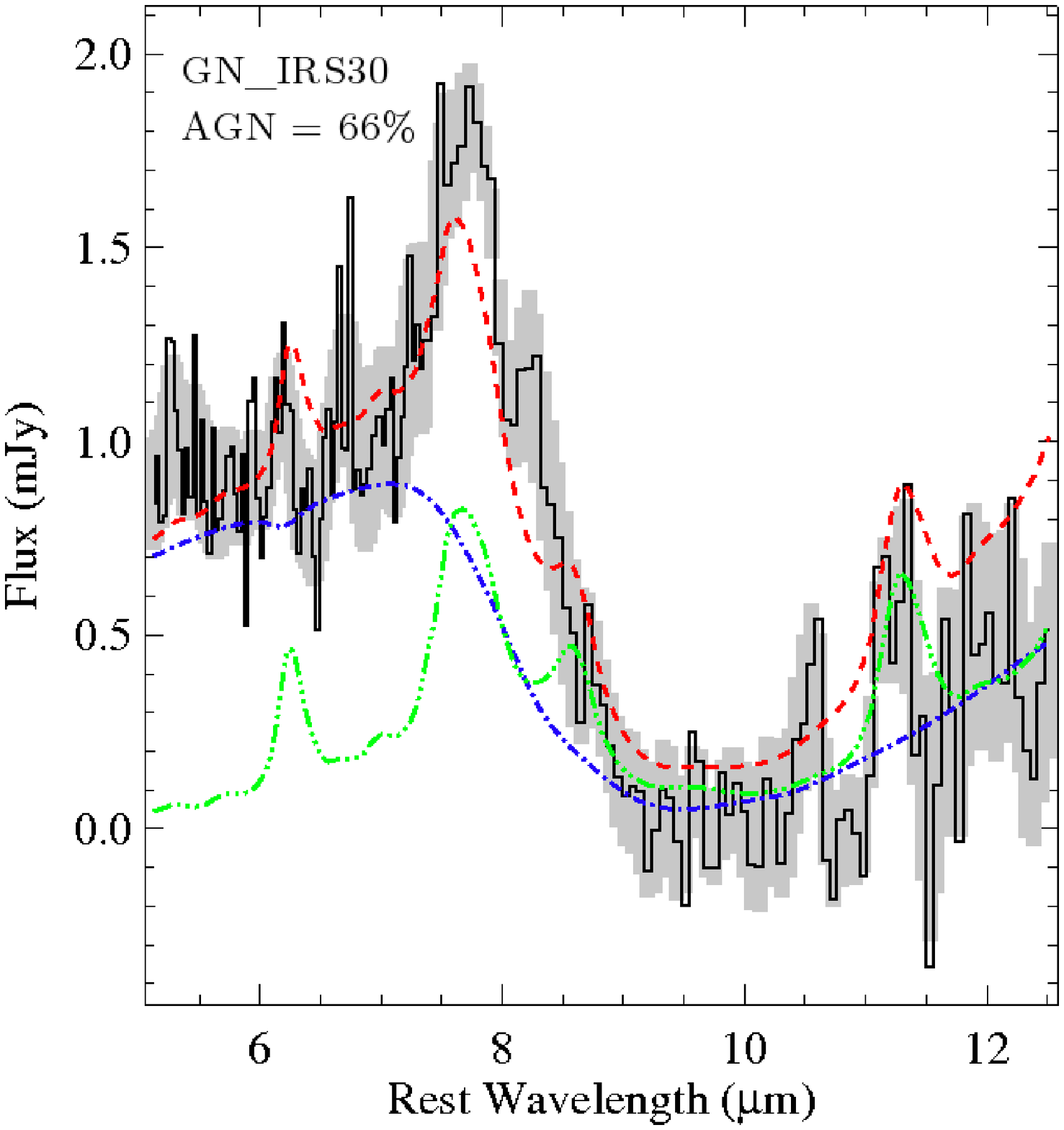}
\caption{Mid-IR spectra of GN\_IRS26 (left), GN\_IRS12 (middle), and GN\_IRS30 (right). The red dashed curves show the best fit SED which is made up of an extincted power law component (blue dashed line) and a starburst template (green dashed line). GN\_IRS26 has strong PAH features indicative of star formation activity, while GN\_IRS12 and GN\_IRS30 are dominated by the power-law component, indicating
the presence of an AGN. GN\_IRS30 shows a strong 9.7$\,\mu$m Si absorption feature on top of the continuum component. The relative contributions of each component to the mid-IR luminosity determines whether a galaxy is dominated by AGN or star formation activity in the mid-IR. \label{sed_ex}
}
\end{center}
\end{figure*}

We perform spectral decomposition of the mid-IR spectrum (5--12$\,\mu$m rest frame) for each source in order to disentangle the AGN and SF components. We follow the technique outlined in detail in \citet{pope2008a} which we summarize here. We fit the individual spectra with a
model comprised of three components: (1) the star formation component is represented by either the local starburst composite of \citet{brandl2006} or simply the mid-IR spectrum of the prototypical starburst M~82 \citep{schreiber2003} -- with the SNR, wavelength coverage, and spectral resolution of our high redshift spectra, both give equally good fits to the SF component of our galaxies; 
(2) the AGN component is determined by fitting a pure power-law with the slope and normalization
as free parameters; (3) an extinction curve from the \citet{draine2003} dust models is applied to the AGN component. The extinction curve is not monotonic in wavelength
and contains silicate absorption features, the most notable for our wavelength range being at 9.7$\,\mu$m. The local starburst composite and the M~82 template already contain some intrinsic extinction. We tested applying additional extinction to the SF component beyond that inherent in the templates and found this to be negligible for all sources. We fit all three components simultaneously and integrate under the starburst spectrum and power-law continuum to determine the fraction of the mid-IR luminosity ($\sim5$--12$\,\mu$m depending on the redshift of the source) from SF and AGN activity, respectively. For each source, we quantify the strength of the AGN in terms of the percentage of the total mid-IR luminosity coming from the power-law continuum component. Based on this mid-IR spectral decomposition, we find that 38 (25\%) out of our sample of 151 galaxies are dominated ($\ge$~50\%) in the mid-IR by an AGN. Figure \ref{sed_ex} shows examples of the best fit models (red dashed line) for
an SF-dominated galaxy and two AGN-dominated galaxies, with and without prominent 9.7$\,\mu$m extinction. The extincted power law component is shown by the blue dashed line, and the PAH template is the green dashed line.

To more thoroughly compare the mid-IR spectral properties and far-IR SEDs within our sample, we divide our galaxies into four sub-samples based on the results of the mid-IR spectral decomposition. First, each galaxy is classified as
either SF- or AGN-dominated, based on having $<$~50\% or $>$~50\% AGN contribution to the mid-IR luminosity, respectively. We further divide the SF galaxies into two bins: $z\sim1$ ($z<1.5$); and $z\sim2$ ($z>1.5$). For AGN, the mid-IR spectral features have been predicted to reveal the shape of the torus surrounding the AGN. A clumpy torus
produces a power-law spectrum and possibly weak silicate absorption, while the presence of strong silicate absorption suggests a thick obscuring torus \citep{levenson2007,sirocky2008}. We therefore classify the AGN according to the shape of their mid-IR spectrum; those with measurable 9.7$\,\mu$m silicate absorption (hereafter referred to as silicate AGN), and those without (hereafter referred 
to as featureless AGN), which we have classified by eye. We have a much smaller number of AGN sources, so separating further according to redshift would not produce a meaningful sample with which to determine the average properties. We are unable to classify four AGN sources as they lack spectral coverage in the relevant range (9--10$\,$\micron), so we are incapable 
of  determining whether they exhibit silicate absorption. We refer to these as unclassifiable AGN in the relevant figures. Our four sub-samples are listed in Table \ref{basictbl}, along with their 
median redshifts and 8, 24, and 100$\,\mu$m flux densities.

\begin{figure}
\centering
\includegraphics[scale=0.7]{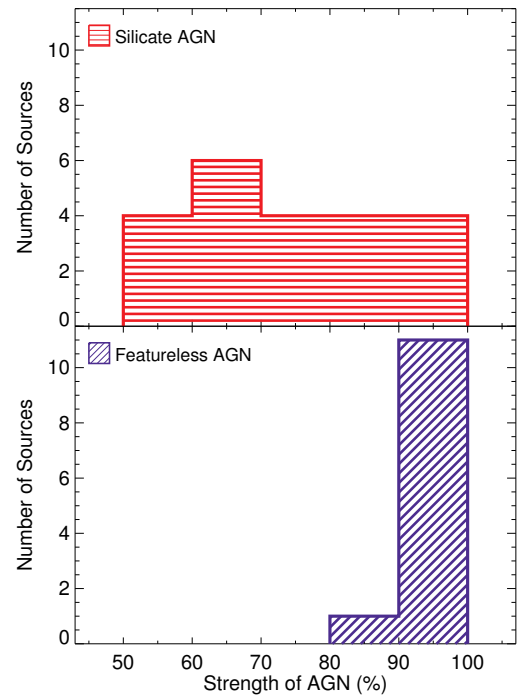}
\caption{Sources whose mid-IR spectrum is dominated by an AGN continuum are further divided according to whether they exhibit significant 9.7$\,\mu$m silicate absorption. The majority of the sources completely dominated by AGN emission ($>$~95\%) exhibit a featureless power-law spectrum (bottom panel), whereas sources with a silicate absorption feature are more uniformly distributed according to the strength of the AGN in the mid-IR (top panel). \label{agnhist}}
\end{figure}

While the majority of our sources that are classified as SF-dominated, based on the mid-IR spectra, have a negligible ($<20$\%) contribution from an AGN, the AGN-dominated sources exhibit varying degrees of concurrent SF activity.
Figure \ref{agnhist} shows the distribution of mid-IR AGN fraction for the silicate (top panel) and featureless AGN (bottom panel). The featureless AGN (lacking silicate absorption) have a very strong AGN continuum, accounting for 80--100\% of the mid-IR emission, whereas the silicate AGN have a more uniform distribution of AGN fraction, with some silicate AGN also having weak PAH features.
The broad range of AGN fraction for the silicate sources is partly a reflection of the difficulty in disentangling PAH features from the 9.7$\,\mu$m silicate absorption feature in low SNR spectra.

\subsection{Spectroscopic Redshifts}
\label{sec:red}
We determine redshifts for the majority of our sample by fitting the positions of the main PAH features \citep[see][]{pope2008a}. In the case of featureless spectra (only $9/151$ sources), we adopt available optical spectroscopic redshifts (\citealp{szokoly2004, barger2008, popesso2009}; Stern et al. in preparation), with the exception of
one source 
for which we only have a photometric redshift. More details on the optical and IRS-based redshifts of individual sources in our sample are given in Pope et al.~(in preparation). 
We compare our redshifts derived from the IRS spectrum with the optical spectroscopic redshifts using the parameter $\Delta z/(1+z_{\rm IRS})$.
We find the mean value to be 0.008 with a standard deviation of 0.07. Therefore, we conclude that our IRS redshifts are good to $\sim$ 0.07. 

\begin{figure}
\includegraphics[width=2.5in]{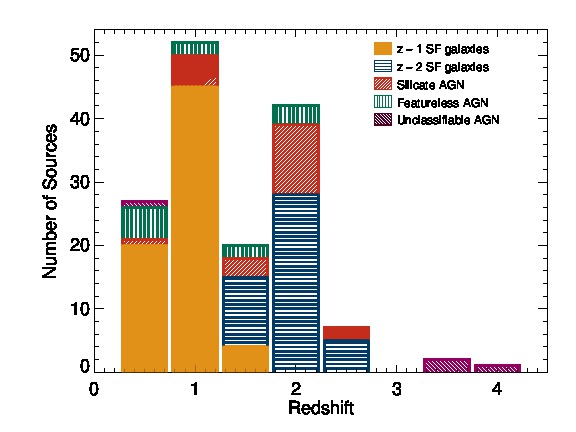}
\caption{Redshift distribution of sources. The distribution appears to be bimodal with peaks at $z\sim1$ and $z\sim2$ reflecting the 24$\,\mu$m selection criterion. At redshifts 1 and 2, the observed frame 24$\,\mu$m band corresponds to the rest-frame 12$\,\mu$m and 8$\,\mu$m band, respectively. In SF galaxies, those wavelengths are dominated by prominent PAH features if they lie at these particular redshifts, causing them to appear more
luminous at 24$\,\mu$m. The sources have been separated according to four sub-samples 
based on the mechanism driving the mid-IR luminosity. Our three highest redshift sources are all AGN, though due to lack of coverage at 9.7$\,\mu$m, we are unable to classify them. \label{zdist}}
\end{figure}

The redshift distribution of our sources is illustrated in Figure~\ref{zdist}. The redshift distribution exhibits a bimodality
with two peaks at redshifts of around 1 and 2. The bimodality of the redshift distribution is a result of the 24$\,\mu$m selection \citep[e.g.,][]{desai2008}. At $z\sim2$, the 24$\,\mu$m observed frame
flux density corresponds to the 8$\,\mu$m flux density in the source's rest frame. Galaxies with intense star formation will have an enhanced 8$\,\mu$m flux density due to prominent PAH complexes at 7.7 and 8.6$\,\mu$m.
Similarly at $z\sim1$, observed frame wavelength 24$\,\mu$m corresponds to a rest frame wavelength of 12$\,\mu$m. The strong PAH emission features at 11.3$\,\mu$m and 12.7$\,\mu$m,
present in SF galaxies, will boost the observed 24$\,\mu$m flux density. Fig. \ref{zdist} shows that our sources at the highest redshifts are all dominated by an AGN in the mid-IR. 
While the featureless AGN are distributed roughly equally across our redshift range, there is a slight dearth of silicate AGN at $z\sim1.5$ due to the 9.7$\,\mu$m absorption feature falling into the 24$\,\mu$m band.

In the sections that follow, we use the median redshift for each subsample (listed in Table \ref{basictbl}) to compare general properties of the subsamples and investigate evolutionary connections between the populations. 
For our SF galaxies, Fig. \ref{zdist} illustrates that the median redshifts of 1.0 and 1.9, respectively, are accurate for each subsample. The median redshift for the
silicate AGN is 1.9, and again, the distribution is peaked around this value. However, the featureless AGN have a fairly uniform redshift distribution with the mean and median
redshifts both giving $z=1.2$. It is important to keep in mind that as we did not separate the AGN based on redshift, we are not claiming that our subsamples are representative of all
AGN at each of the median redshifts.

\subsection{Stellar Masses}
\label{sec:mass}
We use optical and near-IR photometry to estimate the galaxy stellar masses in our sample, and compute the median stellar mass for each sub-sample (see Table \ref{lumtbl}). The stellar masses for the current
sample are part of a larger catalog of stellar masses, photometric redshifts, and multi-wavelength data for galaxies in the GOODS fields,
a detailed description of which will be given in a separate paper (Pannella et al., in preparation).
In the following, and for the sake of completeness, we will briefly describe the sources of the photometry used to calculate the stellar masses and the procedures used.

We have built a PSF-matched multi-wavelength catalog with 10 passbands from\,{\em U} to 4.5$\,\mu$m. The optical, near-IR, and IRAC data used to calculate the stellar masses in GOODS-N are presented in \citet{capak2004, wang2010, lin2012}; and Dickinson 
et al. (in preparation), respectively. A \,$K_{\rm{s}}$-band selected catalog has been built using SExtractor \citep{bertin1996}
in dual image mode, closely following the procedure described in \citet{pannella2009b} and \citet{strazzullo2010}.
For the ECDFS sources we have used photometry from both the GOODS-MUSIC \citep{santini2009} and the GOODS-MUSYC \citep{cardamone2010} surveys to estimate stellar masses. 

Stellar masses are estimated for both fields using the SED fitting code described in detail in \citet{drory2004,drory2009}.
We parametrize the possible star formation histories by a two-component model,
consisting of a main, smooth component described by an exponentially declining star formation rate ($\psi(t) \propto \exp(−t/\tau)$),
linearly combined with a secondary burst of star formation. The main 
component time-scale $\tau$ varies in $\in$ [0.1, 20] Gyr,
and its metallicity is fixed to solar. The age of the main component,\, $t$, is allowed to vary between 0.01$\,$Gyr and the age of the Universe at each 
object's redshift. The secondary burst of star formation, which cannot contain more than 10\% of the galaxy's total stellar mass, is modelled as a 100 Myr old constant star formation rate episode of 
solar metallicity. We adopt a \citet{salpeter1955} initial mass function for both components, with lower and upper mass cutoffs of 0.1 and 100 M$_\odot$, respectively.
Adopting the \citet{calzetti2000} extinction law, both the main 
component and the burst are allowed to exhibit a variable amount of attenuation by dust with \, $A_V$ $\in$ [0,~1.5] and [0,~2] for the main component and the burst, respectively.
We confirmed our masses by also fitting without the secondary burst component and find consistent results.

A significant fraction ($\sim$ 66\%) of our sources that are AGN-dominated in the mid-IR do not display a prominent stellar bump. For these objects, we do not expect the photometry to diverge from
the best-fit results until after 1.5$\,\mu$m, at which point the stellar bump becomes diluted. Based on our fitting method, we find that this should produce a bad fit rather than a biased one. We test
our fitting method for power law sources on an individual basis by removing the IRAC photometry to ensure we were only fitting wavelengths bluer than the 1.6$\,\micron$ bump. The results are consistent
with the results produced when the IRAC data is included. A recent study by \citet{hainline2012} found that for IRAC power-law sources, the stellar mass can be a factor of 1.4 higher when an AGN component is not accounted for in the SED modeling. In this study, we use the stellar masses to calculate specific star formation rates to assess whether our sources lie on the main sequence (see \S \ref{sec:MS}). The possible bias of 1.4 introduced by not 
including an AGN component in the stellar fitting will not greatly affect the discussion  in that section given the spread in the main sequence.

In Section \ref{sec:discuss}, we compare with stellar masses presented in \citet{elbaz2011}. Our stellar masses were calculated using the stellar population synthesis models from \citet{bruzual2003},
while \citet{elbaz2011} use templates from PEGASE.2 \citep{fioc1999}. As a result, our masses are a factor of $\sim2$ lower \citep[see also][]{bell2011}.

\section{Composite Spectral Energy Distributions}
\label{sec:sed}
As described in detail above, we divide our full sample into the four sub-samples listed in Table \ref{basictbl}. 
For each sub-sample, we create a composite SED from 0.3 to 600$\,\mu$m rest-frame
by combining data from: ground-based near-IR; {\em Spitzer} IRAC, IRS, and MIPS (24$\,$\micron, 70$\,$\micron); {\em Herschel} PACS and
SPIRE; and ground-based 870$\,\mu$m and 1.15$\,$mm. 

For sources lacking a detection at the {\it Herschel} wavelengths, we extract a measurement of the flux density and associated uncertainty for each source directly from our images using the prior position at 24$\,\mu$m. This is not just an upper limit but rather an actual best estimate of the flux of the source even if it not formally detected. We take measurements from the image to avoid biasing our SEDs high in the submm by including only detections. We do not take a measurement when a source looks too blended on the image; we have rejected 26 sources from our composite SEDs due to blending in the {\em Herschel} images.

Since spectroscopic redshifts are known for all sources in our sample, we begin by shifting all spectra and photometry to the rest frame. We then examine the full IR SED of each galaxy; sources which lack a sufficient amount of data in the far-IR/submm ($<$~3 measurements beyond a rest wavelength of 25$\,\mu$m) are not used in creating the composite SEDs to avoid biasing our results. 
By requiring sources to have measurements (not necessarily detections) at longer wavelengths, we have only rejected a limited number of sources (26, or 17\%) which are severely blended with other sources in the large SPIRE beams. Our rejected sources have the same redshift distribution as our full sub-samples, ensuring that we have not introduced a redshift bias.

When combining the spectra and photometry for the composite SEDs, we normalize the data for each source to the mid-IR spectrum over the wavelength range 6.4 to 7.5$\,\mu$m (with the exception of the featureless AGN),
chosen because it is free of prominent PAH features, and because the majority of our sources have spectral coverage in this range.
Any sources with exceptionally noisy data over this range (such that we are unable to reliably decompose the AGN and SF components) or without continuous data from 6.4~--~7.5$\,\mu$m are excluded from the composite SEDs. The featureless AGN have
mid-IR spectra that are free of PAH and silicate features, so we choose to normalize over a longer wavelength range, $6.4-11.4\,\mu$m.

For clarity, we now state exactly how many sources are rejected from the creation of the composites in each subsample. In the $z\sim1$ SF galaxies subsample, we begin with
69 sources; we reject 13 due to blending in the far-IR and 17 that lack coverage in the mid-IR normalization range; this leaves 39 sources. For the $z\sim2$ SF galaxies, we begin with
44 sources; we reject 7 sources due to far-IR blending, 5 sources that lack coverage in the mid-IR normalization range, and 2 sources with exceptionally noisy mid-IR spectra; this leaves 30 sources. For the silicate AGN, we begin with 22 sources; we reject 4 due to blending in the far-IR and 1 because it lacks spectral coverage in the mid-IR normalization range, leaving us
with 17 sources. Finally, in the case of the featureless AGN, we initially have 12 sources, from which we reject two that are blended in the far-IR and one because it lacks coverage in the mid-IR normalization range; we therefore have 9 sources which we use to create the composite. The number of sources comprising each composite are listed in Table \ref{basictbl}.
The remaining 95 sources are representative of our full IRS sample:  82\% have $S_{24} > 200\,\mu$Jy, and 60\% have $S_{24} > 300\,\mu$Jy. For all four subsamples, the redshift distributions are the same for the full subsample and the limited subsample used to create the composite. The specific sources we have rejected in each subsample
are indicated in Appendix A.

We determine the median $L_\nu$ for each source over this wavelength range. Within each sub-sample, we also determine the median $L_\nu$ for all sources,
and we normalize the rest-frame data of each source to the median $L_\nu$ of that sub-sample in order to preserve the intrinsic average luminosity of each sub-sample. 

In creating the composite SEDs, we treat the near/mid-IR data separately from the far-IR/submm data. This is because in the far-IR the individual photometric uncertainties dominate the noise, whereas in the near/mid-IR the scatter between different sources in each composite is a larger source of error.  
After normalization, we average the near-IR and mid-IR data by determining the median $L_\nu$ and wavelength in bin sizes $\ge$~0.1$\,\mu$m (larger than the spectral resolution
of the IRS spectra),
chosen so that each bin is well-populated ($>$~10 data points).
We bootstrap the sample to estimate the errors on our composite in the mid-IR and near-IR, since in this regime the uncertainty in the composites is dominated by the scatter between different sources. For each sub-sample, we randomly draw sources with replacement and recalculate the normalized median SED 10,000 times. Because we normalize in
the range $6.4-7.5\,\mu$m, the resulting SEDs exhibit little scatter in and around these wavelengths.

The far-IR data has two sources of error: the instrument noise and the confusion noise. We account for both in our fitting by adding the instrument noise in quadrature with the confusion noise (from \citet{nguyen2010}) for
each flux measurement.
We fit all of the normalized far-IR ($>18\,\mu$m) data in each sub-sample with a two temperature component modified blackbody function of the form 
\begin{equation}
M_\nu=a_1\times \nu^{\beta}\times B_\nu(T_{\rm warm})+a_2\times \nu^{\beta}\times B_\nu(T_{\rm cold})
\end{equation}
where $B_\nu$ is the Planck function. We keep the emissivity fixed at $\beta=1.5$.
Since the photometric uncertainty is large in the far-IR, we use a Monte Carlo simulation to determine the parameters and errors of the blackbody model. We randomly sample each data point within its error and refit the model 10,000 times. We fit for the normalization factors $a_1$ and $a_2$ together with $T_{\rm warm}$ and $T_{\rm cold}$, the temperatures of the warm and cold dust components, simultaneously using a $\chi ^2$ minimization technique. The final parameters for each composite are the mean values determined through the Monte Carlo simulations, and the errors on each parameter are the
standard deviation. We then propagate the standard 
deviation on each parameter, along with the correlations between parameters, which we find to be much less than the standard deviations, to obtain the error bars on the far-IR SED model. The degeneracies between the
parameters are illustrated in Appendix B. We do find a degree of degeneracy between the warm and cold dust temperatures, but the temperature ranges spanned by each component do not overlap, giving us confidence
that we are measuring two distinct temperatures.

\begin{figure*}
\centering
\includegraphics[width=7in]{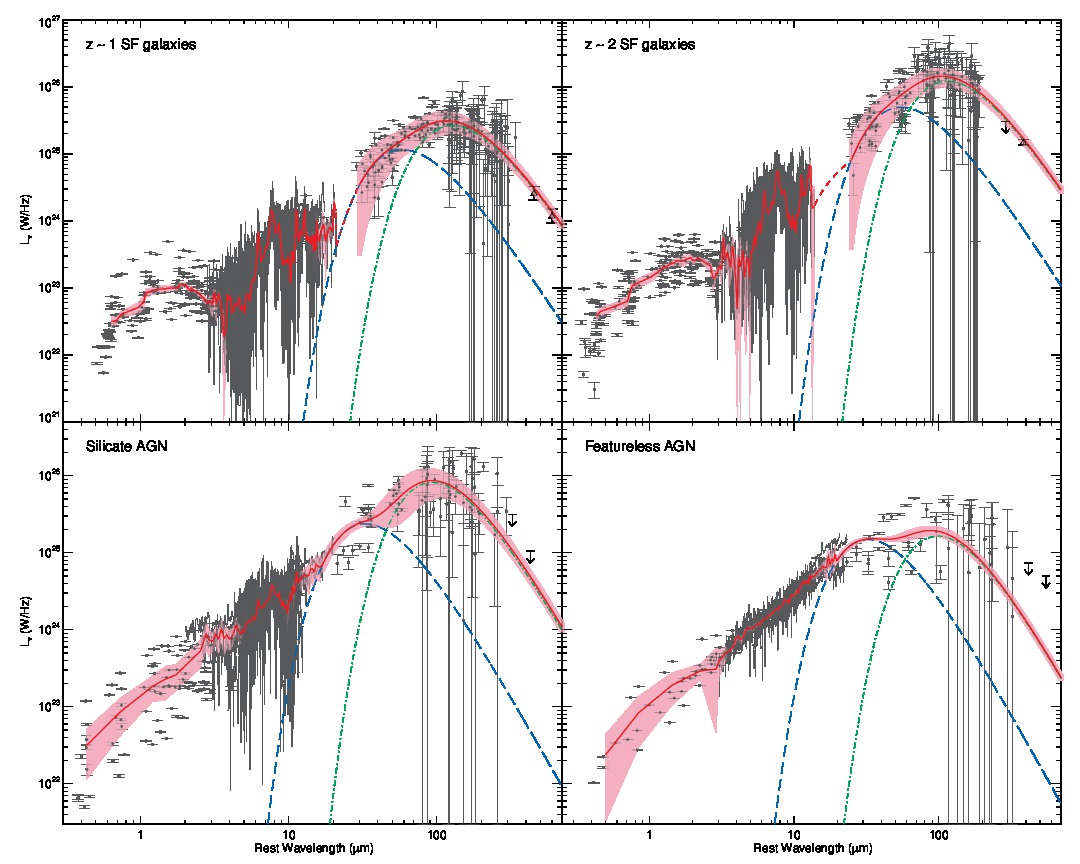}
\caption{Composite SEDs for the mid-IR spectral sub-samples listed in Table \ref{basictbl}. In each panel we show photometric (black points)
and spectroscopic data (black lines) for the sub-sample in the rest frame with our composite SED overplotted in red and the uncertainty of the composite shown as the shaded region.
The errors on the far-IR data points include both the instrument noise and confusion noise.
The near-IR and mid-IR data were averaged to obtain the composite while
the far-IR data were fit with a two temperature modified blackbody model. The blue and green dashed lines illustrate the modified blackbody curves
of the individual warm and cold dust components, respectively.
The open triangles in the upper panels are the stacked (sub)mm detections at observed frame 870$\,\mu$m and 1.15$\,$mm. In the absence of stacked detections, we plot the 3$\sigma$ upper limit. In all cases, the (sub)mm detections and upper limits are consistent with our composites. 
\label{fourpanel}}
\end{figure*}

We settled on a two temperature model after finding very poor fits (high reduced $\chi^2$ values, a factor of $\sim$ 6 higher) to a single component modified blackbody. The introduction of a second dust component produces a
significantly better fit, with lower $\chi^2$ values.
Since the individual photometric uncertainties dominate the noise in the far-IR composites, each data point is weighted by its associated variance in our fits. In the case of the SF
SEDs, both sets of data lack photometry from $\sim 18 - 30\,\mu$m. The warm modified blackbody overlaps with the mid-IR stacking in the $z\sim1$ SF SED, so we use this
modified blackbody to fill in the gap in the SED we were unable to fit. For the $z\sim2$ SF SEDs, the warm modified blackbody does not overlap with the stacked mid-IR spectrum,
so we linearly interpolate between the stacked photometry and fitted far-IR photometry to fill in the area with no photometry. Both of these regions are indicated with the dashed lines
in Figure \ref{fourpanel}. 
We verify the far-IR fits by calculating the medians in differential bin sizes for the far-IR data using the same techniques described for the near-IR data, and in all cases, our model fits match the binned medians.
 Our composite SEDs for each of the four sub-samples are shown in Fig. \ref{fourpanel}. 

We use longer ground-based (sub)mm measurements to verify the accuracy of our composite SEDs. Most of our sources are not individually detected in the LABOCA and AzTEC+MAMBO maps, and therefore we stack all sources in each sub-sample to obtain an average flux density at 870$\,\mu$m and 1.15$\,$mm, respectively. We follow the (sub)mm stacking method outlined in \citet{pope2008b}. 
The stacking of the SF sub-samples results in several $>4\sigma$ detections which we plot on the composite SEDs in Fig. \ref{fourpanel} as the triangles. When the stacking results in a non-detection, we plot a 3$\sigma$ upper limit.
The (sub)mm detections lie right on the composite SEDs for the two SF sub-samples, confirming our modified blackbody model fits. For the AGN composites, the upper limits from stacking are also consistent with our composite SEDs.

In Section \ref{sec:modparam} we discuss the key features of our new high redshift SEDs. Given that we report the average luminosity, redshift, and stellar mass of the sources that go into these templates, they may be useful for fitting other high redshift sources, though it is important to bear in mind that we created the composites from sources that are
bright in the mid-IR ($S_{24} > 200\,\mu$Jy). We make these new empirical high redshift SED templates publicly available\footnote{http://www.astro.umass.edu/$\sim$pope/Kirkpatrick2012/}.

\section{Features of Composite SEDs}
\label{sec:modparam}

\subsection{Mid-IR spectral features}
\label{sec:SEDdecomp}
We perform mid-IR spectral decomposition on our composites as we did for the individual galaxies (see Section \ref{sec:decomp}).
Since the majority of the sources comprising both composite SF SEDs are completely dominated by star formation,
we naturally find that our SF composites both have negligible ($<10\%$) AGN contribution in the mid-IR.
The featureless AGN composite SED is 100\% dominated in the mid-IR by an AGN power-law component.
The silicate AGN SED is arguably the most interesting, since it contains both weak PAH features and a strong continuum component.
Upon decomposition, we find the silicate AGN composite to have an 84\%
AGN contribution in the mid-IR. Both the SF SEDs and the featureless AGN SED have negligible silicate absorption at 9.7$\,\mu$m,
while $\tau_{9.7}$~=~0.4 for the silicate AGN composite SED. The silicate AGN SED has a power-law slope of $\alpha = 1.5$ from 4 -- 15$\,\mu$m, and the
featureless AGN SED has a power-law slope of $\alpha = 1.6$ in the same wavelength range. Both numbers are consistent with what is derived from local AGN in \citet{mullaney2011}.

\begin{figure}
\plotone{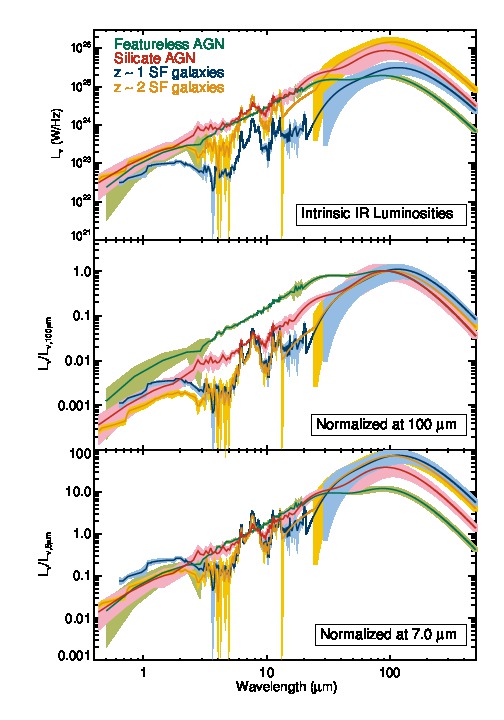}
\caption{Comparison of the composite SEDs created from the four sub-samples of galaxies. {\em Top panel~--}~No additional normalization has been applied; the SEDs represent the average intrinsic luminosities of sources in each sub-sample. The shaded regions in the mid-IR and near-IR 
correspond to the 1$\sigma$ spread of sources around the median luminosities. In the far-IR, the shaded regions are computed by propagating the errors on each parameter present in the blackbody
fitting routine. The $z\sim2$ SF SED is the most luminous at $\lambda >$~100$\,\mu$m, while the featureless AGN SED is the weakest at the longest wavelengths. The 
AGN SEDs are the most luminous at 20$\,\mu$m where the warm dust component dominates the spectrum.
{\em Middle panel --} Composite SEDs have been normalized at 100$\,\mu$m, allowing for more direct comparison of the near and mid-IR. No variation of PAH features is seen between SF composite
SEDs. The featureless AGN SED has the highest mid-IR to far-IR ratio.
{\em Bottom panel --} Composite SEDs have been normalized at 7$\,\mu$m. The weak PAH features of the silicate AGN composite are visible in direct comparison to the SF SEDs. \label{composites}}
\end{figure}

\subsection{Dust Temperature and Infrared Luminosity}
Our full composite SEDs are shown together in Figure \ref{composites}. All SED composites peak at approximately the same wavelength in the far-IR ($\sim100\,\mu$m,) which implies that each sub-sample has cold dust at approximately
the same temperature. 
However, the full SEDs clearly show a difference in the relative amounts of warmer dust, which we quantify by fitting a two temperature modified blackbody model (see Table \ref{temptbl} for the dust
temperatures for each sub-sample). A two temperature model can be physically understood as follows: the cold dust ($\gtrsim 100\,\micron$) is due to the diffuse interstellar medium, heated by the
underlying stellar population, whereas the warm dust originates in SF regions and traces the younger stellar population \citep[e.g.,][]{bendo2012}. In the case of an AGN, as the AGN becomes luminous
the dust surrounding it heats up and can radiate in the mid-IR and far-IR. Therefore, we might expect to see warmer dust in our AGN SEDs than in our SF SEDs. Naturally, a multi-temperature
model is even more ideal, as the AGN is not the only heating source for the dust. In an AGN still actively forming stars, there should a be a third dust temperature that arises from the SF regions.
Unfortunately, even for our large amount of photometric data in the far-IR, fitting more than two dust temperatures becomes degenerate and produces poor $\chi^2$ values. Therefore, we use two 
temperatures with the assumption that these represent the dominant sources of dust radiation, but there are likely more sources of heating that we are unable to separately quantify.

For the AGN composites, there is a degree of uncertainty as to where the power-law spectrum ends and the warm dust component begins. Due to limited spectroscopic data,
we are only able to stack spectra below $\sim\,18\,\mu$m, and we fit our dust temperatures using only far-IR data (above 18$\,\mu$m). Recent work has suggested that the
power-law portion of the AGN spectrum continues to 20$\,\mu$m, flattens, and then falls off after 40$\,\mu$m \citep{mullaney2011}. In this case, assigning a single temperature to this portion of the
spectrum is misleading, since it is composed of many dust temperatures. However, our primary motivation is to compare our AGN composite SEDs directly with our SF SEDs, so we choose
to fit a single blackbody to the warm dust emission around $\sim20 - 60\,\mu$m. Since the AGN is also radiating into the mid-IR power-law portion of the spectrum, the IR luminosity we derive from the warm dust component only will be a lower limit on the AGN emission.

There is a known degeneracy between the dust emissivity, $\beta$, and temperature, such that if we fix $\beta$ to a higher value, it will produce lower temperatures. We chose to
fix $\beta$ rather than fit it along with the temperatures, but it is important to keep the degeneracy of the two parameters in mind when interpreting the temperatures. There is evidence that $\beta$
is universal on galaxy-wide scales \citep{dunne2001}, and as we fix it to the same value for every sub-sample, we do not expect that there is any bias created between the sub-samples in this study.

From our fitting we find that the two AGN composites have cold dust temperatures of $\sim$~33~K, while the SF galaxies are only slightly colder at $\sim$~27~K.
However, the AGN possess a significantly hotter warm dust component ($\sim$~99~K) in comparison to the warm dust component for the SF galaxies ($\sim$~57~K). This makes sense if the warm dust emission is predominantly heated by the AGN as opposed to star formation in dense HII regions.

To quantify the relative contributions of the hot and cold dust components, we calculate $L_{\rm IR}$ for each component separately by integrating over each modified blackbody model
(see Table \ref{lumtbl}).
We compare $L_{\rm IR}^{\rm warm}$ and $L_{\rm IR}^{\rm cold}$ to the total IR luminosity (integrated over 8 to 1000$\,\mu$m), $L_{\rm IR}$, to determine the fraction of $L_{\rm IR}$ attributed to each dust component. We then
calculate a luminosity-weighted average temperature by multiplying each luminosity fraction
by the respective temperature and combining to obtain an effective temperature, $T_{\rm eff}$ (Table \ref{temptbl}). $T_{\rm eff}$ for the AGN SEDs is more than 20~K hotter than $T_{\rm eff}$ for
the SF galaxies, demonstrating the bolometric significance of the warm dust component in AGN-dominated sources. The two SF composites have almost the same $T_{\rm eff}$, indicating little evolution
in $T_{\rm dust}$ with redshift in SF galaxies.

\begin{deluxetable*}{lccc}
\tablecaption{Dust temperatures derived for each sub-sample. \label{temptbl}}
\tablehead{ \colhead{} & \colhead{Effective} & \colhead{$T$} & \colhead{$T$} \\
\colhead{Sub-sample} & \colhead{Temperature \tablenotemark{a}} & \colhead{(warm dust) \tablenotemark{b}}& \colhead{(cold dust) \tablenotemark{b}}}
\startdata
$z\sim1$ SF galaxies 	& 37 $\pm$ 3 K &  \, 55 $\pm$ 6 K & 25 $\pm$ 2 K \\
$z\sim2$ SF galaxies 	& 40 $\pm$ 3 K &  \, 59 $\pm$ 5 K & 28 $\pm$ 2 K \\
Silicate AGN 			& 58 $\pm$ 2 K &  \, 98 $\pm$ 2 K & 35 $\pm$ 3 K \\
Featureless AGN 		& 65 $\pm$ 2 K & 100 $\pm$ 1 K & 33 $\pm$ 1 K
\enddata
\tablecomments{See Table \ref{lumtbl} for IR luminosities.}
\tablenotetext{a}{Calculated from a luminosity-weighted average of the warm and cold dust components. See text for details.}
\tablenotetext{b}{Temperatures are the mean values determined from Monte Carlo simulations (see \S \ref{sec:sed}), with the errors being the respective standard deviations}
\end{deluxetable*}

The $L_{\rm IR}$ values computed by integrating each composite SED from 8 to 1000$\,\mu$m are listed in Table \ref{lumtbl}; differences between the sub-samples are largely due to our flux-limited sample. 
$L_{\rm IR}$ for the composite SED created from $z\sim2$ SF galaxies is five times greater than $L_{\rm IR}$ for the $z\sim1$ SF composite; these luminosities qualify the $z\sim1$ and $z\sim2$ SF SEDs to be LIRGs and ULIRGs, respectively.
The silicate AGN possess a notably larger $L_{\rm IR}$ ($1.7\times 10^{12}\, {\rm L}_{\odot}$) than the featureless AGN ($L_{\rm IR}=8\times 10^{11}\, {\rm L}_{\odot}$), though this is likely due to the
different redshift distributions of the two sub-samples and not any intrinsic evolution of $L_{\rm IR}$.
The luminosities of the individual dust components (listed in Table \ref{lumtbl}) quantify the relative importance of each dust component in the different SEDs. 
For the featureless AGN, the warm dust component is more luminous than the cold dust component by a factor of $>2$. Though the warm dust does not dominate the far-IR SED of any other sub-sample, it does
contribute at a level of $\sim50\%$ of the total IR luminosity. 

This is an important result, since it demonstrates that single temperature blackbody models or templates derived using only longer wavelength SPIRE data will miss the importance of this warm dust component. For example, \citet{smith2012} derive composite SEDs for $z<0.5$ galaxies primarily using SPIRE data, which we can compare with our composite SEDs. The median SED of galaxies with $L_{\rm dust} > 10^{11}\, {\rm L}_{\odot}$ from \citet{smith2012} is consistent with our SF composite SEDs in the mid-IR and beyond the peak in the submm, but it is a factor of $\sim2$ lower than our SF composites in the wavelength range 20 -- 60$\,\mu$m, where the warm dust dominates. The discrepancy arises from the fact
that \citet{smith2012} do not possess PACS data for the majority of their sources and so they may miss a significant warm dust component. 
This comparison underscores the importance of full photometric coverage when modeling far-IR SEDs. The optimal strategy is to combine deep PACS and SPIRE data to obtain a full census of the dust emission.

If, instead of a two temperature modified blackbody, we fit our full suite of far-IR data above $\sim 20\,\mu$m with a one temperature modified blackbody, we get a poor $\chi^2$ fit
(reduced $\chi^2$ a factor of $\sim$ 6 higher than the reduced $\chi^2$ values for the 2 temperature fits, due to the fact that data $> 70\,\mu$m is not fit well),  and we calculate an $L_{\rm IR}$ that is $\sim 30\%$ lower for our SF composites and silicate AGN composite.
For the featureless AGN composite, using a one temperature modified blackbody lowers the $L_{\rm IR}$ by $\sim20\%$. We also derive temperatures that are $\sim20$ K hotter than the cold dust temperatures (see Table \ref{temptbl}) for all four sub-samples, with the temperatures of the SF composites comparable to that of the AGN composites. This
underscores the importance of the two temperature approach, since data points in the range 24 -- 100$\,\mu$m can bias any one-temperature fit to a warmer dust temperature and lower $L_{\rm IR}$.

On the other hand, when photometry spanning the range 24 -- 100$\,\mu$m are missing, the dust emission will likely be biased to colder temperatures and lower luminosities.
For our SF composites, only using the cold dust for calculations lowers both the $L_{\rm IR}$ and star formation rates by a factor of $\sim 2$. The biggest discrepancy arises
from the featureless AGN, where the warm dust is more important than for the other SEDs. There, the warm dust accounts for ~60\% of the total IR luminosity. When dealing with individual galaxies,
often there are not enough data points to unambiguously fit both the warm and cold dust components. As not accounting for any warm dust may significantly bias results, it is perhaps best to use templates which account for both important dust components as observed in high redshift galaxies. 

\subsection{Overall SED shape}

In the top panel of Fig. \ref{composites}, which compares the intrinsic luminosities of the sources going into each composite, the AGN SEDs have more emission than the SF SEDs in the range 15 to 30$\,\mu$m. At $z\sim2$, the MIPS 70$\,\mu$m passband covers this wavelength range. In fact,
for our high redshift sources ($z>1.5$), 64\% of those detected at 70$\,\mu$m are AGN dominated. For our lower redshift sources ($z<1.5$), only 19\% of 70$\,\mu$m-detected sources are
dominated by an AGN in the mid-IR. At the peak of the SED, the most luminous sources are the $z\sim2$ SF galaxies and silicate AGN, reflecting their higher total IR luminosities listed Table \ref{lumtbl}.

We normalize the composite SEDs to 100 and 7$\,\mu$m in the middle and bottom panels of Fig. \ref{composites}, respectively, to allow a more direct comparison of the mid-IR
spectral features. The mid-IR slopes of the AGN SEDs are remarkably similar, despite the absence/presence of features, but the featureless AGN have a much larger mid-IR to far-IR ratio. In fact, the
featureless AGN SED effectively flattens out at $\lambda >$~20$\,\mu$m. Both AGN SEDs lack a visible stellar bump at 1.6$\,\mu$m, indicating that the rest-frame near-IR emission is also dominated by the AGN. 
The SF and silicate AGN SEDs all exhibit PAH features, though they are much weaker for the silicate AGN SED.

In the middle and bottom panels of Fig. \ref{composites}, we can directly compare our two SF composite SEDs to look for any evolution in SED shape with redshift and/or $L_{\rm IR}$. 
The PAH features of the two SF SEDs are nearly identical in strength
and shape. There appears to be no evolution of PAH features between LIRGs at $z\sim1$ and ULIRGs at $z\sim2$. The far-IR part of the SF SEDs are also remarkably similar in shape, as quantified in the average dust temperatures (Table \ref{temptbl}). 
The stellar bump at 1.6$\,\mu$m is more prominent relative to the IR emission for the $z\sim1$ SF SED than for the $z\sim2$ SF SED, reflecting the higher average specific star formation rate of galaxies at $z\sim2$ compared to $z\sim1$, which we discuss in the next section.

\subsection{Star Formation Rates}
\label{sec:SFR}
We calculate the star formation rate (SFR) for each composite SED according to the formula
\begin{equation}
\left (\frac{\rm SFR}{\mbox{M}_\odot \mbox{ yr}^{-1}}\right)=1.72 \times 10^{-10}\left(\frac{ L^{\rm SF}_{\rm IR}}{\mbox{L}_\odot}\right)
\end{equation}
which assumes a Salpeter IMF \citep{salpeter1955} and continuous star formation \citep{kennicutt1998}. 
Note that we only want to use the portion of $L_{\rm IR}$ that is heated by SF activity when using this equation, otherwise we will over-predict the SFR. 
For the SF composite SEDs, we assume that both the cold and the warm ($\sim60\,$K) dust components are heated by star formation, since there is no evidence of AGN activity in the composites based on the mid-IR spectrum.

We calculate the fraction of the AGN SED that can be accounted for by the SF SED template when both are normalized at $200\,\mu$m, beyond which all the dust emission is presumed to be due to SF.
We calculate the $L_{\rm IR}$ of the normalized SEDs and take the ratio of $L_{\rm IR}$ for the $z\sim1$ SF SED and featureless AGN SED, and for the $z\sim2$ SF SED and silicate AGN SED.
We use the SF composite SED that most closely matches the total $L_{\rm IR}$ of the respective AGN composite when determining the $L_{\rm IR}$ ratios. We find that SF activity accounts for 56\% of the
$L_{\rm IR}$ for the silicate AGN SED and 21\% of the featureless AGN SED. We therefore scale the $L_{\rm IR}$ by 56\% and 21\% for the silicate and featureless AGN, respectively, when calculating
the SFR. 

Unsurprisingly, our $z\sim2$ SF composite SED has the highest SFR: 344\,M$_\odot$ yr$^{-1}$. The $z\sim1$ SF composite SED has an SFR of 73\,M$_\odot$ yr$^{-1}$. The silicate AGN SED has an average SFR of 159\,M$_\odot$ yr$^{-1}$, whereas the featureless AGN SED has a much lower average SFR of 28\,M$_\odot$ yr$^{-1}$. If we instead assume that all of the $L_{\rm IR}$ in the AGN composites is heated by SF, we would over-predict the SFR by a factor of $\sim2 - 4$.

As a consistency check, for the two AGN SEDs, we conservatively attribute the warm dust at 100$\,$K to the AGN and the cold dust to the SF activity and recalculate the SFR using 
$L_{\rm IR}^{\rm cold}$ instead of the scaled $L_{\rm IR}$; naturally, this conservative estimate should provide a lower limit on the actual SFR,
since there is likely some portion of warmer dust emission that arises from SF. Using just the cold dust component, we find the same SFR of $28\,{\rm M}_\odot\, {\rm yr}^{-1}$ for the featureless AGN, and an SFR of
$141\,{\rm M}_\odot\, {\rm yr}^{-1}$ for the silicate AGN, which are very similar to the SFRs we calculated above, so most of the SF activity in each AGN composite is accounted for by the cold dust component.
  
There is a tight correlation between stellar mass and SFR which evolves with redshift \citep{daddi2007,dunne2009,pannella2009a,elbaz2011,lin2012}, and as such, an interesting comparison is given by the specific star formation rates (sSFR), defined as: sSFR = SFR/$M_{\ast}$. We list the median stellar mass (see Section \ref{sec:mass}) and sSFR of each composite SED in Table \ref{lumtbl}.
For the SF composites, which have comparable median stellar masses, we see an increase in the sSFR from $z\sim1$ to $z\sim2$, consistent with what is observed for all SF galaxies \citep[e.g.,][]{elbaz2011}. 
The silicate AGN and $z\sim2$ SF SEDs have comparable stellar masses and redshifts, but the sSFR for the $z\sim2$ SF SED is more than a factor of 2 higher. Similarly, the featureless AGN composite has a higher stellar mass and $\sim$ 4 times lower sSFR than the $z\sim1$ SF composite. 
As we will discuss in Section \ref{sec:discuss}, these values are consistent with a scenario in which the AGN sources are in a later phase in the evolution of IR luminous galaxies, after the active period of star formation has been quenched.

\begin{deluxetable*}{lccccccccc}
\tablecolumns{10}
\tablecaption{Luminosities and other parameters derived for each sub-sample. \label{lumtbl}}
\tablehead{\colhead{} & \colhead{$L_{\rm IR}$(8--1000$\,\mu$m)} & \colhead{$ L_{\rm IR}^{\rm warm}$ \tablenotemark{a}}
& \colhead{$L_{\rm IR}^{\rm cold}$ \tablenotemark{a}} & \colhead{\underline{$L_{\rm IR}^{\rm cold}$}} &
\colhead{$L_{8}$\tablenotemark{b}} & \colhead{\underline{ $L^{\rm total}_{\rm IR}$ }} & \colhead{$M_{\ast}$} & \colhead{SFR} & \colhead{sSFR}\\
\colhead{Sub-sample} & \colhead{(10$^{12}$ L$_\odot$)} & \colhead{(10$^{11}$ L$_\odot$)} & \colhead{(10$^{11}$ L$_\odot$)} & 
\colhead{$L^{\rm total}_{\rm IR}$} & \colhead{(10$^{11}$ L$_\odot$)} & \colhead{$L_{8}$} & \colhead{($10^{11}$ M$_\odot$)} &\colhead{(M$_\odot$ yr$^{-1}$)} & \colhead{(Gyr$^{-1}$)}}
\startdata
$z\sim1$ SF galaxies & 0.42 $\pm$ 0.17 & 2.0 $\pm$ 0.5 & \ 2.0 $\pm$ 0.6  & 0.46 & 0.65 & 6.57 & 0.74 &  73 $\pm$ 29 & 0.99 \\
$z\sim2$ SF galaxies & 2.00 $\pm$ 0.71 & 8.7 $\pm$ 1.9 & 10.2 $\pm$ 2.1 & 0.51 & 2.53 & 7.92 & 1.17 & 344 $\pm$ 122 & 2.94 \\
Silicate AGN		  & 1.65 $\pm$ 0.54 & 7.0 $\pm$ 0.5 & \ 8.5 $\pm$ 2.4 & 0.50 & 2.60 & 6.35 & 1.15 & {\em 159 $\pm$ 52}\,\tablenotemark{c} & 1.38 \\
Featureless AGN		  & 0.76 $\pm$ 0.07 & 4.4 $\pm$ 0.2 & \ 1.6  $\pm$ 0.3 & 0.21 & 1.94 & 3.94 & 1.23 & {\em 28 $\pm$ 3}\,\tablenotemark{c} & 0.23
\enddata
\tablenotetext{a}{Determined by integrating under the modified blackbody curve for the warm and cold components separately.}
\tablenotetext{b}{Measured by integrating under the IRAC 8$\,\mu$m filter for each composite SED.}
\tablenotetext{c}{Calculated using a scaled $L^{\rm SF}_{\rm IR}$ for the AGN SEDs (shown in italics to distinguish).}
\end{deluxetable*}

\subsection{Comparison of New High z SEDs to Local Templates}
\label{sec:compare}
Prior to the era of {\em Herschel}, far-IR (40 -- 200$\,\mu$m) data sampling the peak of the SED were rare, particularly for high redshift sources. Estimates of bolometric luminosity were performed by fitting locally derived
templates to mid-IR and/or submm data and extrapolating to the far-IR. With our wealth of far-IR data from {\em Herschel}, we are in a position to compare the commonly applied locally derived
templates with our new empirical high-redshift SEDs.

\begin{figure}
\plotone{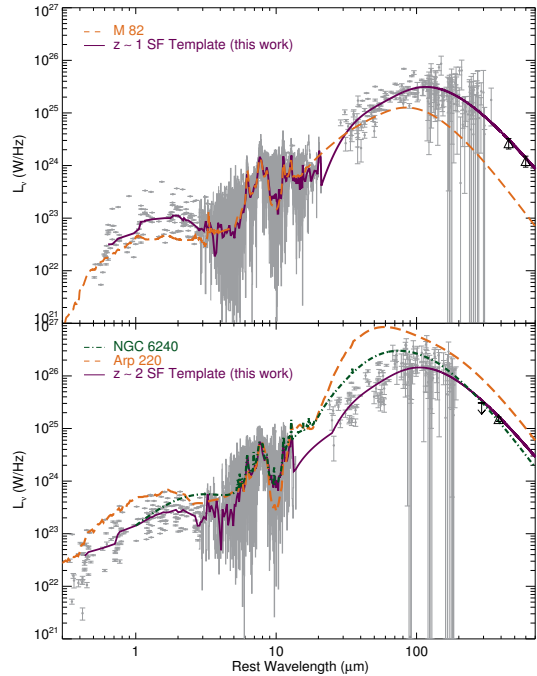}
\caption{Comparison of our sources and composites with local SEDs, overplotted on the photometric
(gray points) and spectroscopic (gray lines) data for a given SF sample. {\em Top --} $z\sim1$ SF composite SED compared to the SED of M~82 (normalized at 7.0$\,\mu$m). M~82 is in excellent agreement with our $z\sim1$ composite SED in the mid-IR, but is weighted towards warmer dust temperatures and underpredicts the luminosity in the far-IR/submm. {\em Bottom --} $z\sim2$ SF composite SED compared to the SEDs of Arp~220 and NGC~6240 (both normalized at 7.0$\,\mu$m). The local and high $z$ ULIRG SEDs show fairly good agreement in the range 5~--~11$\,\mu$m, but both local SEDs are too luminous in the far-IR, particularly at the shortest wavelengths, indicating more warm dust. In addition, Arp~220 has stronger silicate absorption than our composite SED. \label{sb_local_temp}}
\end{figure}

In Figure \ref{sb_local_temp} we compare our SF composite SEDs to the SEDs of local galaxies often used as standard templates in the literature.
These local templates come from combining all known data on known sources including available IRAS (12, 25, 60, 100$\,\mu$m), {\em Spitzer} (24, 70, 160$\,\mu$m), and SCUBA (850$\,\mu$m) photometry as 
well as mid-IR spectroscopy \citep[e.g.,][]{schreiber2003,armus2007}. The local templates lack photometry from 160 -- 850$\,\mu$m, making it difficult to constrain the cold dust emission.

In the top panel of Fig. \ref{sb_local_temp}, we compare our $z\sim1$ composite SF SED to the SED of local starburst M~82. Though M~82 has a lower $L_{\rm IR}$ ($\sim 4 \times 10^{10}\,{\rm L}_\odot$) than our $z\sim1
$ SED, it is worthwhile to compare the two SEDs, since our mid-IR classification is based on the M~82 template. We normalize M~82 to our composite SED at 7.0$\,\mu$m. Our $z\sim1$ composite
SED almost exactly reproduces the shape and strength of the PAH features of M~82, affirming the reliability of comparing the mid-IR PAH features of this local galaxy with the mid-IR spectra of high 
redshift sources. However, M~82 has less far-IR emission, and it is weighted towards warmer dust than our $z\sim1$ SF composite, indicating that any mid-IR similarities do not carry over into the far-
IR.

The bottom panel of Fig. \ref{sb_local_temp} shows our $z\sim2$ composite SF SED compared to two local prototypical (U)LIRGs, Arp~220 ($L_{\rm IR} = 1.4 \times 10^{12}\,{\rm L}_\odot$) and NGC~6240 ($L_{\rm IR} = 6.3 \times 10^{11}\,{\rm L}_\odot$), both normalized to our composite SED at 7.0$\,\mu$m. 
While the mid-IR spectrum of NGC~6240 is very close to that of our high redshift composite, Arp~220 shows substantially more silicate absorption. 
The prominent absorption feature at 9.7$\,\mu$m is possibly attributable to AGN activity in Arp~220 \citep[e.g.,][]{armus2007}, while our $z\sim2$ composite SF SED has a negligible AGN contribution. 
Neither of these local (U)LIRGs match our $z\sim2$ SED in the near or far-IR; 
specifically, Arp~220 and NGC~6240 peak at shorter wavelengths in the far-IR, indicating much more warm dust emission. It is possible the difference in warm dust emission arises from the fact that
local ULIRGs are merger-induced, triggering AGN activity, while our sample of high redshift ULIRGS may not be (see Section \ref{sec:MS}).
These local ULIRG templates differ significantly from our $z\sim2$ ULIRG composite, and in fact, there are not even any individual sources that go into the composite that fit the local ULIRGs from 20 -- 80$\,\mu$m (see the gray points showing all galaxies that go into the composite). This illustrates the difficulty of applying such local templates to high redshift systems.

\begin{figure}
\centering
\includegraphics[width=3in]{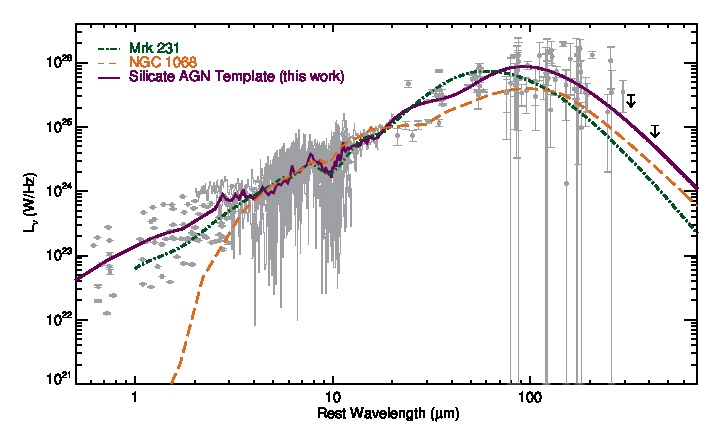}
\caption{Comparison of our silicate AGN composite SED with SED templates for the local AGN Mrk~231 and NGC~1068.  The templates are overplotted on the photometric
(gray points) and spectroscopic (gray lines) data for the silicate AGN sample. Both Mrk~231 and NGC~1068 have
been normalized to the silicate AGN SED at 7.0$\,\mu$m. Mrk~231 peaks at shorter wavelengths in the far-IR than our template, while NGC~1068 falls off sharply below 3$\,\mu$m. \label{agn_local_temp}}
\end{figure}

We compare our silicate AGN composite SED with the SEDs of local AGN Mrk~231 ($L_{\rm IR} = 3.2 \times 10^{12}\,{\rm L}_\odot$) and NGC~1068 ($L_{\rm IR} = 1.6 \times 10^{11}\,{\rm L}_\odot$) in Figure \ref{agn_local_temp}. Both Mrk~231 and NGC~1068 have
been normalized to our composite at 7.0$\,\mu$m. Our composite and Mrk~231 appear to have a similar 9.7$\,\mu$m silicate absorption feature and a similar slope in the mid-IR, but Mrk~231 peaks at shorter far-IR wavelengths than our composite SED, indicating more warm dust. NGC~1068 looks fairly similar in shape in the far-IR to our composite, though it is intrinsically less luminous. Furthermore,
NGC~1068, a local Compton-thick AGN, has less silicate absorption in the mid-IR and falls off sharply below 3$\,\mu$m, differing from both our high redshift SED and Mrk~231. 

\begin{figure}
\plotone{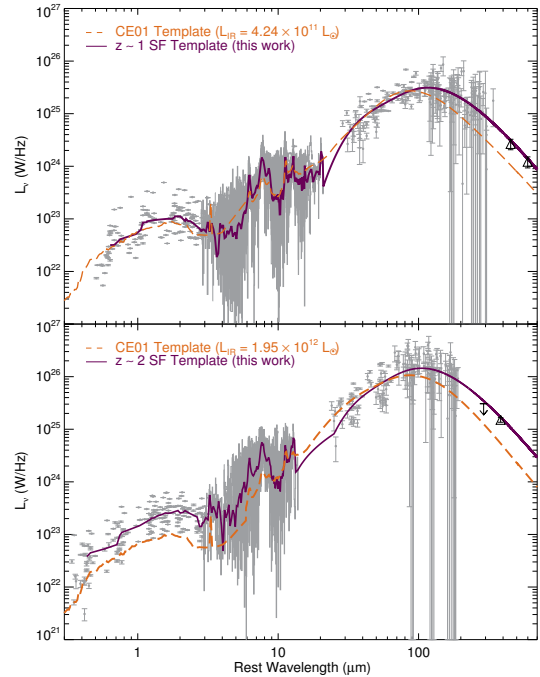}
\caption{Comparison of our $z\sim1$ (top) and $z\sim2$ (bottom) SF composite SEDs with the locally-derived CE01 templates of the same luminosity. The templates are overplotted on the photometric
(gray points) and spectroscopic (gray lines) data for a given SF sample.
For the $z\sim1$ SF galaxies, the CE01 template traces the composite SED in the near and mid-IR but differs in the far-IR. For the $z\sim2$ SF galaxies, the CE01 template
fails to reproduce any portion of the composite SED. \label{ce01}}
\end{figure}

With the emerging abundance of {\em Herschel} photometry, recent studies have begun to demonstrate the unsuitability of locally-derived templates in fitting high redshift
galaxies using longer wavelength PACS and SPIRE data \citep{dannerbauer2010,elbaz2010,elbaz2011,nordon2010}. These previous results fit templates from \citet[CE01]{chary2001} templates to individual
galaxies with far-IR photometric data points, which each have significant uncertainty. In this study, we are able to compare the locally-derived CE01 templates to templates derived from high-redshift data for a statistical sample of galaxies. 
We plot the CE01 
templates with the $L_{\rm IR}$ corresponding to each of our SF composite SEDs in Figure \ref{ce01}, with no renormalization.
For the $z\sim1$ SF composite (top panel), the CE01 template reproduces fairly well the near and mid-IR portions of the composite SED.
However, it peaks at slightly shorter far-IR wavelengths and has less cold dust emitting at (sub)mm wavelengths. 
For the $z\sim2$ SF composite (bottom panel),
the CE01 template fails to reproduce any portion of our $z\sim2$ composite SED. 
The mid-IR spectrum of the CE01 template has a strong continuum component and weaker PAH features than we observe in the average $z\sim2$ ULIRG of similar luminosity. Again the far-IR emission peaks at shorter wavelengths, corresponding to warmer dust emission, than our $z\sim2$ SF SED. The stacked (sub)mm points (plotted as the triangles) are also inconsistent with the CE01 templates, confirming
that our high redshift sample has more cold dust emission than the CE01 templates. 

Overall, we find that most local templates fail to accurately reproduce both the dust temperatures and mid-IR spectral features of our high redshift composite SEDs. 
It is possible that things will be more consistent once {\it Herschel} data are incorporated to create more complete SEDs of local galaxies and better constrain the cold dust emission.
Otherwise, this is evidence that the SEDs of IR luminous galaxies evolve with redshift. The disparity of the locally derived templates from our high redshift composites illustrates the
need for caution when applying local templates to high redshift galaxies.

\section{Discussion}
\label{sec:discuss}

In Sections \ref{sec:sed} and \ref{sec:modparam}, we presented our high redshift composite SEDs and outlined key differences and similarities between the SF- and AGN-dominated IR luminous galaxies. We also compared our new high redshift SED templates to those of local galaxies of similar luminosities and found them to be significantly different. We now discuss what these results might tell us about the evolution of IR luminous galaxies.

\subsection{Quantifying the AGN Component}
\label{sec:xray}
We begin by exploring the X-ray and mid-IR properties of the AGN comprising our composite SEDs.
Though an in-depth X-ray analysis of our AGN sample is beyond the scope of this paper, we briefly discuss two important X-ray parameters commonly used to interpret AGN: X-ray detection fraction and X-ray luminosity. Our entire sample of AGN sources (38) have a 58\% X-ray detection fraction. When we classify our sources into silicate and featureless AGN, only 
$9/22$ (41\%) silicate AGN are individually detected in the deep {\em Chandra} X-ray imaging, whereas $11/12$ (92\%) featureless AGN are detected in the X-ray. The four unclassifiable AGN have a detection fraction of 50\%.

We calculate the intrinsic X-ray luminosity, $L_{\rm 2-10\,keV}$, using equation (4) of \citet{mullaney2011}.
\begin{equation}
\log\left(\frac{L^{\rm AGN}_{\rm IR}}{10^{43}\, {\rm erg\, s}^{-1}}\right) = 0.53\, +\, 1.11 \log\left(\frac{L_{\rm 2-10\,keV}}{10^{43}\, {\rm erg\, s}^{-1}}\right)
\end{equation}
To disentangle the luminosity due to the AGN from the host galaxy, we follow the same method used to calculated the SFR (see \S \ref{sec:SFR}). We normalize all composite SEDs to 200$\,\mu$m and calculate the ratio of the normalized $L_{\rm IR}$ for the $z\sim1$ SF SED and featureless AGN SED, and for the $z\sim2$ SF SED and silicate AGN SED.
We find that AGN activity accounts for 44\% of the
$L_{\rm IR}$ for the silicate AGN SED and 79\% of the featureless AGN SED, and accordingly we scale each respective $L_{\rm IR}$ by these amounts when calculating
$L_{\rm 2-10\, keV}$. This gives an intrinsic X-ray luminosity of $L_{\rm 2-10\, keV} = 5.30 \times 10^{44}\, {\rm erg\, s}^{-1}$ for the silicate AGN and $L_{\rm2-10\, kev} = 4.46 \times 10^{44}\, {\rm erg\, s}^{-1}$ for the
featureless AGN.

\begin{figure}
\centering
\includegraphics[width=3in]{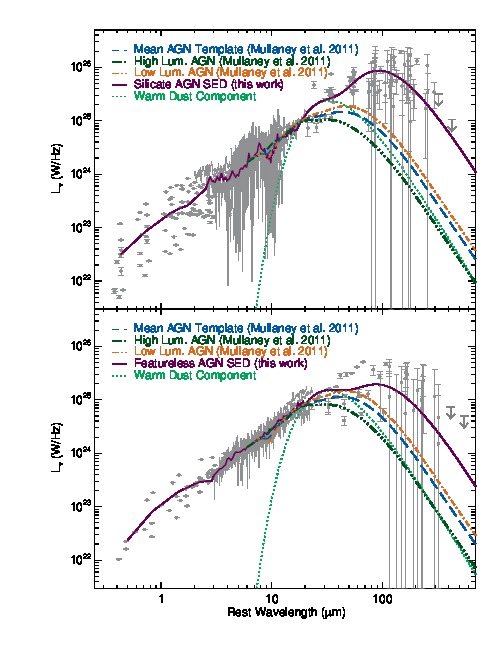}
\caption{We compare our composite AGN SEDs (solid lines) with the average AGN SEDs derived in \citet{mullaney2011}, normalized at $7\,\mu$m. The blue dashed line shows the mean AGN SED, while the green and orange dot-dashed
lines show the high and low luminosity AGN SEDs, respectively, which have been created using the SEDs of local AGN. We overplot the warm dust component for each of our composite SEDs as the dotted line. The warm dust emission peaks at the
same wavelength as the high luminosity AGN SEDs, which is consistent with the high $L_{2-10\,{\rm kev}}$ values we derive for our composites.\label{mullaney}}
\end{figure}

We look for evolution between our high redshift AGN and local AGN by comparing to the sample of moderate luminosity ($L_{\rm 2-10\,keV} \sim 10^{42}-10^{44}\,{\rm erg\,s}^{-1}$) AGN in \citet{mullaney2011}. By cross-matching the {\em Swift}-BAT sample \citep{tueller2008} of X-ray AGN with the {\em Spitzer}-IRS archive, \citet{mullaney2011} identified a sample of nearby (i.e., $z<0.1$) galaxies whose mid-infrared spectra are
strongly AGN-dominated. By carefully subtracting any contribution from the host galaxy at far-IR wavelengths, they were able to empirically define the infrared emission due solely to the AGN in these local galaxies and derive an average, intrinsic AGN SED at $6-100\,\mu$m. In Figure \ref{mullaney}, we plot the mean AGN SED and the high and
low luminosity AGN SEDs from the local population along with our high redshift AGN composite SEDs. The high and low luminosity AGN SEDs were derived by
averaging the SEDs of
$\log(L_{\rm 2-10\,keV}) > 42.9$ and $\log(L_{\rm 2-10\,keV}) <42.9$ AGN, respectively. The low luminosity AGN template and the mean AGN template turn over around $40-50\,\mu$m whereas the high luminosity AGN template turns over at 32.0\,$\mu$m. In both of our high $z$ AGN composites, the warm dust component, which we attribute to the AGN, turns over at approximately the same wavelength (29\,$\mu$m) as the high luminosity local AGN template.

The mid-IR spectral decomposition of our AGN composite SEDs resulted in an 84\% contribution of the AGN to the mid-IR luminosity for the silicate AGN SED and a 100\% contribution for the featureless AGN
SED (see \S \ref{sec:SEDdecomp}). If we make the simplifying assumption that the warm temperature component is due entirely to the AGN and the cold component to the host galaxy, then we can
quantify the contribution of the AGN to the total IR luminosity by subtracting the contribution of $L_{\rm IR}^{\rm cold}$ from $L_{\rm IR}$ (see Table \ref{lumtbl}).
For the silicate AGN SED, we find an IR contribution of 50\%, and for the featureless AGN, a contribution of 79\%, in good agreement with what we find when we remove the host component by scaling the 
SF composite SEDs. Both SEDs exhibit a lower overall contribution from the AGN component to $L_{\rm IR}$ than that indicated by the mid-IR alone, since the SF component dominates at the longest IR wavelengths. 
As a check on the validity of our simple assumptions about the two temperature components, we also compare the $L^{\rm AGN}_{\rm IR}$ that we derive from the \citet{mullaney2011} mean AGN templates (Fig. \ref{mullaney}). Fitting the \citet{mullaney2011} mean AGN templates to the mid-IR of our AGN SEDs, we find an AGN
contribution to $L_{\rm IR}$ of 62\% and 83\% for the silicate and featureless AGN, respectively. These results are remarkably consistent with our modified blackbody fits, which is reassuring. With our blackbody fits, we are able to attach physical parameters, namely dust temperatures, to each population which eases comparisons between our samples and with other galaxy populations at low and high redshift.

\subsection{Evolution of infrared luminous galaxies}

In the local Universe, all ULIRGs and most LIRGs are observed to be caught in the act of a major merger \citep{sanders1996}, where the interaction is the trigger for the intense infrared luminosity
\citep{murphy1996,bushouse2002}. A popular scenario for linking IR luminous galaxies dominated by SF and AGN activity is outlined in \citet{sanders1988}. During the early stages of a merger 
of two disk galaxies a dusty starburst is triggered, but as the black holes merge together and are fed by the disrupted gas,
an AGN begins to dominate the IR emission and destroys or blows away the dust 
and gas, quenching the star formation.
Eventually the AGN runs out of fuel, and the galaxy settles down as a massive elliptical galaxy. This scenario is a plausible explanation for local ULIRGs.

At $z\sim1$ -- 3, the situation is different. While some high redshift ULIRGs are major mergers \citep[e.g.,][]{engel2010}, many high redshift ULIRGs show no evidence for any merger or interaction, from 
studies of their morphology or dynamics \citep[e.g.,][]{daddi2010,tacconi2010}. The higher gas fractions and longer gas consumption timescales suggest that the intense star formation in these galaxies 
is fueled continuously, perhaps through cold gas streams \citep[e.g.,][]{dekel2009}. Several recent studies \citep[e.g.,][]{schawinsky2011,mullaney2012} have found that many moderate luminosity AGN are found in high redshift IR luminous galaxies that are not necessarily undergoing a major merger, suggesting that internal 
instabilities could be the primary mechanism fueling the AGN.

We found significant differences in the IR SED between our mid-IR identified SF and AGN galaxies; specifically, the AGN sources have effective dust temperatures $\sim20\,$K higher,
indicating that the luminous AGN is not only responsible for heating the hot dust in the mid-IR, but also has an impact on the warmer far-IR dust and can even dominate the bolometric luminosity
\citep[e.g. our featureless AGN in Table \ref{lumtbl}, see also][]{diaz2010}.
It is tempting to try to place our SF- and AGN-dominated galaxies into an evolutionary sequence, where the SF galaxies represent an early phase of active star formation, and the AGN galaxies represent 
a later phase once the black hole has been fueled and is able to heat the dust to warmer temperatures and quench the star formation.
This evolutionary sequence may or may not be triggered by a major 
merger at high redshift. Any direct evolutionary comparison can only be tentatively drawn between our $z\sim2$ SF SED and our silicate AGN SED, since these two SEDs are
comparable in $L_{\rm IR}$ and stellar mass, and both are primarily composed of galaxies with a redshift of $\sim2$.
There is some tentative evidence that a major merger produces more warm dust emission than normal SF activity \citep{hayward2012}. 
Our $z\sim2$ SF SED exhibits much less warm dust emission than the local ULIRG SEDs, indicating high redshift ULIRGs are not analogs to local ULIRGs and may follow
a different evolutionary path in which they are normal SF galaxies until the AGN at the center begins to dominate the bolometric luminosity.
Our silicate AGN composite SED is luminous in the X-ray ($L_{\rm 2-10\, keV} \sim 5 \times 10^{44}\, {\rm erg\, s}^{-1}$) and has a lower sSFR than the $z\sim2$ SF composite SED. The AGN properties of our silicate AGN SED are consistent with a 
later evolutionary stage than our $z\sim2$ SF galaxies, regardless of whether the evolution is driven by major mergers or some other process.

\subsubsection{Main sequence of star forming galaxies}
\label{sec:MS}
One way that has been proposed to determine if a high redshift galaxy is undergoing a starburst event triggered by a merger is using the ratio of SFR to stellar mass. Several recent studies have found a tight correlation between SFR and stellar mass which has been called the `main sequence' for star-forming galaxies \citep[e.g.,][]{noeske2007, elbaz2007, daddi2007, magdis2010}.
Furthermore, this main sequence shifts to higher SFR from $z=0$ -- 3, leading to an increase in the average sSFR with redshift \citep[e.g.,][]{elbaz2011}. If galaxies on the main sequence are undergoing continuous star formation then starburst galaxies will lie above the main sequence, their SFRs possibly boosted by a merger or interaction. 

Both of our SF composites and our silicate AGN composite have sSFRs that place them on the sSFR-redshift main sequence defined by Eqns. 13 and 24 in \citet{elbaz2011}
and Eqn. 2 in \citet{pannella2009a}. The featureless AGN, however, lie decidedly below both of these relations, though not in the regime designated as
``quiescent" \citep[$<$ 10\% of the average sSFR at a given redshift,][]{mullaney2012}. Our stellar masses are a factor of $\sim2$ lower than those in \citet{elbaz2011} due to different stellar
population models (see \S \ref{sec:mass}), so if we increase our stellar masses by this amount, our sSFRs will consequently be a factor of 2 lower, placing them even further away from the SB regime. 

\citet{elbaz2011} discovered another tight correlation between the 8$\,\mu$m luminosity, $L_{8\mu m}$, and $L_{\rm IR}$, mimicking the SFR-$M_\ast$ main sequence, and defined an alternate main sequence using the parameter IR8 = $L_{\rm IR}/L_8$. IR8 is calibrated to distinguish between starbursting and main sequence SF galaxies, but it has been shown to be useful when applied to AGN as well.
We calculate the 8$\,\mu$m luminosity of our SF and AGN sources by applying the IRAC 8$\,\mu$m passband to each composite SED in the rest frame. We calculate IR8 = $L_{\rm IR}/L_8$ for each composite and list the values in Table \ref{lumtbl}.
We do not distinguish between main sequence and starburst galaxies for individual sources since we are interested in the average properties of our sub-samples. 
Using a large sample of GOODS-{\em Herschel} galaxies, \citet{elbaz2011} define the main sequence to be IR8 = 2.7 to 7.8 (range is the 68\% dispersion around the mean IR8 value of 4.9).
Interestingly, IR8 for both of our AGN composites fits comfortably within this range.
The $z\sim1$ SF SED has an IR8 of 6.6, which is well within the main sequence, while the $z\sim2$ SF SED, with an IR8 of 7.9,
lies just above the main sequence range. Although most of the galaxies in our $z\sim1$ SF sub-sample are main sequence galaxies, the $z\sim2$ SF sub-sample may contain a mix of both main sequence galaxies and starbursts, according to the IR8 parameter.

During the active stages of a major merger, the AGN produced is likely to be heavily obscured. As our present study does not rely on X-ray detection
to identify AGN, we are not biased towards unobscured AGN. Therefore, if AGN at high redshift primarily reside in (U)LIRGs triggered by a major merger, our main-sequence/starburst classification should
reflect this. As neither the silicate AGN nor featureless AGN lie in the starburst regime, according to the IR8 criterion 
and the sSFR-$z$ relation, our findings are in agreement with recent studies that suggest the majority of AGN at high redshift ($z\sim$ 1 -- 3) reside in normal, main sequence galaxies
\citep[e.g.,][]{lutz2010,mullaney2012,xue2010}. However, two caveats must be taken into consideration. First, a compact starburst and an AGN can, in effect, wash each other out
in the IR8 parameter. The starburst can increase the ratio of the far-IR to the mid-IR, but the AGN will increase the amount of mid-IR luminosity. Together, these effects act to produce
a main-sequence IR8 value. Second, both main sequence criteria may be limited in usefulness when discussing AGN. A major merger may immediately trigger a starburst, but may take
more time to funnel material to the AGN, and this can produce a delay between starburst activity and a luminous AGN on the order of  $\sim10^8$ yrs. It is possible that any signatures of a merger have
disappeared before the galaxy displays strong AGN signatures.

Based on the IR8 parameter, most of our AGN and SF galaxies do not show enhanced IR luminosity and may not be a phase of the popular local ULIRG major merger scenario, although they might still be linked in some other evolutionary sequence (e.g.~involving minor mergers or cold gas accretion). Naturally, our findings are only for the average of all of our high redshift sub-samples of galaxies
and do not necessarily apply on an individual basis.

\begin{figure}
\centering
\includegraphics[width=3in]{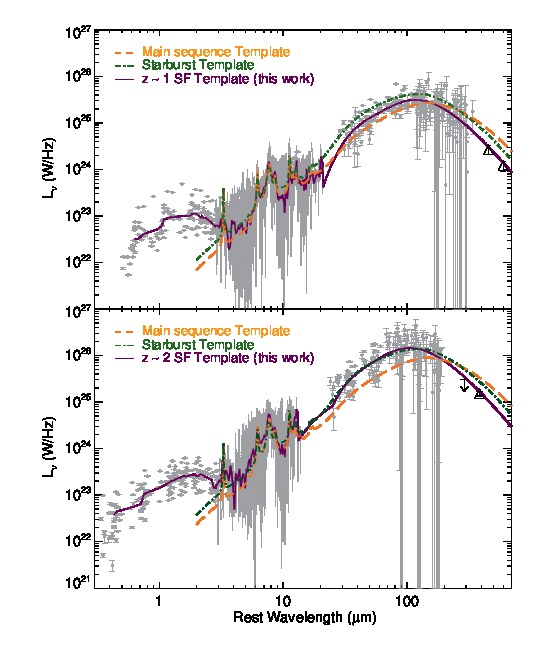}
\caption{Our $z\sim1$ and $z\sim2$ SF SEDs (top and bottom panel, respectively) compared against the main sequence and starburst templates derived in \citet{elbaz2011}.
Both the main sequence and starburst templates overpredict emission at far-IR/submm wavelengths ($\gtrsim 200\,\mu$m).\label{ms_sb}}
\end{figure}

\citet{elbaz2011} present templates for typical main sequence and starbursting galaxies, computed by stacking photometry for main sequence and starburst galaxies at $z<2.5$ from 3 -- 350$\,\mu$m. We compare both of our SF SEDs to the main sequence and starburst templates in Figure \ref{ms_sb}. Upon comparison to our SF templates, we find that both the \citet{elbaz2011} main sequence and starburst templates overpredict the amount of emission at wavelengths greater than 200$\,\mu$m rest-frame exhibited by both our $z\sim 1$ and $z\sim 2$ SF templates, indicating even more cold dust emission. This is possibly due to a difference in the average intrinsic IR luminosity or redshifts for each of these templates. 
In addition, there are differences in the way the templates were created, which could also account for some of the variation in cold dust emission. \citet{elbaz2011} only include the long wavelength SPIRE 
photometry in their fit when there is a detection in the prior catalog whereas we include a SPIRE measurement for every galaxy regardless of if it is a formal detection (see \S \ref{sec:sed}). Furthermore, 
\citet{elbaz2011} fit the far-IR data for individual sources (requiring at least one measurement above 30\, $\mu$m rest wavelength) with the CE01 templates to calculate each $L_{\rm IR}$, and then normalized every source to the same $L_{\rm IR}$ before stacking to create their templates. This can introduce a degree of uncertainty since it requires an extrapolation between the data points.
We normalize out sources in the mid-IR where we have the IRS spectra so we don't need to rely on any templates. The amplitude and shape of our SEDs at these longer wavelengths is validated by our stacked submm measurements (see triangles in 
Fig. \ref{ms_sb}). Most of the mismatch between the \citet{elbaz2011} templates and ours in the submm is likely due to the fact that we are comparing sources at different redshifts and luminosities; the \citet{elbaz2011} SEDs include all galaxies at $z<2.5$ and will be weighted towards galaxies at $z\lesssim1$ with lower average $L_{\rm IR}$, which may indeed be intrinsically cooler, than our SF composite SEDs. As shown earlier in Section \ref{sec:compare}, local templates do not fit the $z\gtrsim1$ galaxies, and we expect the SEDs of star forming galaxies to start to evolve somewhere between $z\sim 1$ and $z\sim 0$. A more in depth study of the SEDs of star forming galaxies at $z<1$ is needed to determine when the SED shapes begin to differ from local templates.

\subsubsection{Obscuration in AGN}

Our AGN sources are clearly split into two sub-samples, those with a measurable silicate feature and those without, and the same is seen in samples of local AGN.  
Silicate absorption has been linked with the most heavily obscured, Compton-thick AGN \citep[e.g.,][]{sales2011, georgantopoulos2011}. Sources with absorption have an optically thick, smooth dust distribution, whereas clumpy dust produces only shallow absorption \citep[e.g.,][]{spoon2007,nenkova2008}, although recently \citet{goulding2012} find that the deep silicate absorption features seen in local Compton-thick AGNs are predominantly caused by extinction due to the host galaxy. According to \citet{levenson2007}, based on
modeling local AGN, the strength of the silicate absorption feature in our silicate AGN SED does not necessarily require the SED to be heavily obscured; nevertheless, there
are significant differences between the silicate AGN and featureless AGN subsamples. Interestingly, in our sample only 41\% of silicate AGN are individually detected in the deep {\em Chandra} X-ray imaging, whereas 92\% of featureless AGN are detected in the X-ray. This supports the idea that the silicate AGN are more heavily obscured, possibly even
Compton-thick \citep{alexander2008}, on an individual basis, while the featureless AGN are less obscured and can have a stronger effect on the host galaxy.
A more detailed study of the individual X-ray properties of our IRS sample is reserved for a future paper. 

Broadly speaking, silicate AGN may represent the transition stage from a star formation dominated galaxy to an unobscured AGN, following the \citet{sanders1988} prescription for creating luminous QSOs. 
For our AGN sources, we assume the mid-IR emission arises from the hot dust torus surrounding the AGN, while the emission at $\gtrsim 60\,\mu$m arises from the cold dust heated by star formation in the host galaxy.
The persistence of significant cold dust observed in our silicate AGN suggests that these nuclei are not yet having a strong effect on the star formation occurring in the host galaxy. 
For an unobscured AGN, the mid IR emission from the warm dust in the torus will dilute the dust emission in the surrounding galaxy, so the estimated amount of cold dust is suppressed or overrun by the warm dust. Currently, a precise picture of AGN feedback has not been proven, but one prominent line of thought is that as the AGN becomes more powerful, the feedback from the AGN is responsible for actively shutting down star formation. AGN feedback could be responsible for the lack of any SF features or silicate absorption in the mid-IR, the suppression of cold dust emission, and the much lower SFR observed in our featureless AGN compared to our silicate AGN. Alternatively, the featureless AGN sources might just be 
more luminous AGN, such that they outshine any of the host galaxy emission, although our inferred X-ray luminosities are inconsistent with this (see \S \ref{sec:xray}). Of course, while 
these scenarios are possible in general, based on the average properties of our sub-samples of galaxies, it is impossible to state that this is the case for every galaxy within each 
sub-sample, since we are probing a range of redshifts and $L_{\rm IR}$. Furthermore, with the level of silicate absorption we see in our templates, it is difficult to distinguish between a 
clumpy and smooth torus geometry. In this case, it is possible that we are simply seeing identical AGN at different viewing angles which produces the different shapes in the spectra. 

While our SEDs are consistent with a scenario in which the featureless AGN are in a later evolutionary stage than the silicate AGN, and are having more of an impact on the host galaxy, evidenced by the lack of mid-IR features, the suppression of cold dust, and the higher X-ray detection fraction, we cannot rule out other explanations for the differences between the SEDs. Detailed modeling of emission from different AGN components that contribute to our composite SEDs is beyond the scope of this paper. Fully disentangling why IR luminous AGN at 
high redshift have noticeably different spectral shapes is a necessary step to producing a complete picture of galaxy evolution.

\subsection{A closer look at our high redshift SEDs}

Looking only at the high redshift SF sources, we might ask whether there is evidence for evolution in the SEDs between $z\sim1$ and $z\sim2$. We showed in Fig. \ref{composites} that when normalized in the mid-IR or far-IR, our two SF composite SEDs are indistinguishable over 3 -- 500$\,\mu$m; the only difference is seen in the near-IR, where the $z\sim1$ SF SED is a factor of a few higher than the $z\sim2$ SF SED. This region of the SED is probing the stellar bump and is a rough proxy for stellar mass. When we compare the average sSFR = SFR/$M_{\ast}$ of our $z\sim1$ and $z\sim2$ SF galaxies, we find an increase of a factor of $\sim2.5$, consistent with the observed increase in the sSFR with redshift \citep[e.g.,][]{elbaz2011}. 
In addition, galaxies with more dust emission (as evidenced by a higher IR luminosity) will have more extinction of the stellar bump. Beyond these two effects, there is no evidence from our analysis 
that the SEDs of IR luminous SF galaxies evolve between $z$ = 1 and 2, though we are comparing LIRGs at $z\sim1$ with ULIRGs at $z\sim2$. The lack of any strong differences between our $z\sim1$ and $z
\sim2$ SF SEDs is especially interesting, since the former are primarily normal main sequence galaxies, while the latter contain a mix of main sequence and starburst galaxies according to the IR8 
parameter. This suggests that mergers either might not significantly alter the distribution of dust and gas in high redshift ULIRGs, or might not be as prevalent as they are in local ULIRGs.
Recently, \citet{diaz2011} analyzed the extended mid-IR emission for a large sample of local LIRGs and ULIRGs, and found that once the central component of the galaxy is removed,
the average SEDs of LIRGs and ULIRGs are all remarkably similar in shape to the local starburst template of \citet{brandl2006}. Their results imply that the differences between LIRG and ULIRG SEDs are
due to processes taking place in the cores. High redshift ULIRGs are more extended than their local counterparts, so the extended emission could be dominating the shape of the SED, causing
little difference between the $z\sim1$ and $z\sim2$ SEDs. 

On the other hand, we do see an evolution in the SEDs of SF galaxies when we compare our high redshift composite SEDs to those of local ULIRGs. The general trend is that we find more of the IR luminosity is coming from colder dust at high redshift, whereas local ULIRGs have more shorter wavelength far-IR emission from warmer dust. This was originally observed for high redshift submillimeter selected galaxies \citep[e.g.,][]{pope2006}, and we now show that it extends to a larger high redshift ULIRG population selected in the mid-IR. 
We argue that this result would not be as robust without modeling the IR emission with multiple dust components since a single dust component would bias the far-IR peak to slightly shorter wavelengths,
and would fail to properly
account for all of the substantial warmer dust.
The presence of more cold dust in high redshift ULIRGs is likely to be in part because they are more extended than local ULIRGs (even major mergers at high redshift are more extended) allowing for a larger fraction of the dust to remain at cooler temperatures \citep{daddi2010,engel2010,ivison2011}. An interesting test of our empirical SED templates would be to compare them to the expected SEDs from radiative transfer models that take into account the larger sizes of high redshift ULIRGs.

\section{Summary}
We have combined deep photometry from 3.6 to 500$\,\mu$m with {\em Spitzer} mid-IR spectroscopy for 151 high redshift ($z>$~0.5) (U)LIRGs in order to explore the relative contribution
of AGN and SF activity to the total IR emission. Our sources bright in the mid-IR (93\% have $S_{24} > 100\,\mu$Jy), and we have shown that they are representative of the full 24\,$\mu$m population above this flux limit. We perform spectral decomposition on the mid-IR spectrum of each source to determine whether the mid-IR luminosity is dominated by
continuum emission from an AGN or PAH emission from ongoing star formation.

Based on the mid-IR spectral decomposition, we divide our sample into 4 sub-samples ($z\sim1$ SF, $z\sim2$ SF, silicate AGN, featureless AGN) and use 95 sources with complete spectral coverage to create composite SEDs over the full IR range for
each sub-sample. We fit a two temperature component modified blackbody model to our composite SEDs and calculate the dust temperatures and IR luminosities. By comparing the properties of the SEDs, we conclude the following:
\begin{enumerate}[(i)]\itemsep0pt
\item We find no significant difference in the SEDs (PAH features, dust temperatures) between SF systems with similar stellar masses at $z\sim1$ (average $L_{\rm IR} = 4.2 \times 10^{11}\, \mbox{L}_\odot$)
and $z\sim2$ (average $L_{\rm IR} = 2.0 \times 10^{12}\, \mbox{L}_\odot$).
\item AGN dominated galaxies with a pure power law spectrum are dominated by a warm ($\sim100\,$K) dust component in the IR, whereas AGN with silicate absorption have a strong warm dust component and  show a significant cold dust component.
\item We find warmer ($\sim$~20~K) effective average dust temperatures for our AGN SEDs than for our SF SEDs.
\item We find that a single temperature model is a poor fit to the data and neglects significant emission at shorter far-IR wavelengths leading to inaccurate $L_{\rm IR}$s and SFRs.
\end{enumerate}

We compare our composite SEDs created from high redshift (U)LIRGs with local templates, including the library of \citet{chary2001}, and find that the local templates differ significantly
from our SEDs, particularly at far-IR wavelengths where ULIRGs at high redshift contain a higher fraction of cool dust than local ULIRGs. The difference in warm dust emission
between local and high redshift SEDs could be attributed to different evolutionary paths. As the local SEDs used in this work are extrapolated from 160 -- 850$\,\mu$m, far-IR {\em Herschel}
photometry is needed for stronger comparisons to be made between the cold dust emission in local and high redshift galaxies.

We use the criteria presented in \citet{elbaz2011}
to assess the `starburstiness' of our sub-samples and find, according to either the IR8 parameter or sSFR-$z$ relation, that most of our sources lie on the main sequence of star formation. 
The properties of our SF and AGN sub-samples are consistent with scenarios which link the evolution of star formation and rapid black hole growth in galaxies at high redshift.
\\

\acknowledgments
We are grateful to the referee for thoughtful comments that improved the clarity of this paper. 
We thank N. Drory for sharing the SED-fitting code used to estimate galaxy stellar masses. 
This work is based in part on observations made with the {\it Herschel Space Observatory}, a European Space Agency Cornerstone Mission with significant participation by NASA, and the {\it Spitzer Space Telescope},
which is operated by the Jet Propulsion Laboratory, California Institute of Technology under a contract with NASA. Support for this work was provided by NASA through an award issued by JPL/Caltech. 
 This work is also based in part on observations obtained with WIRCam, a joint project of CFHT, Taiwan, Korea, Canada, and France, at the Canada-France-Hawaii Telescope (CFHT) which is operated by the National Research Council
(NRC) of Canada, the Institute National des Sciences de l’Univers of the Centre National de la Recherche Scientifique of France, and the University of Hawaii.\\

\renewcommand{\thefigure}{A-\arabic{figure}}
\setcounter{figure}{0}

\appendix

\section{A. Far-IR Photometry of our IRS sources}
In the table below, we present mid-IR and far-IR photometry for our 151 IRS sources. We list MIPS 24 and 70\,$\mu$m, PACS 100 and 160\,$\mu$m, and SPIRE 250, 350, 500\,$\mu$m photometry. We list the results of
the mid-IR spectral decomposition for each source as a percentage of the mid-IR luminosity that is due to the power-law component. We also list the redshifts we calculated from the IRS spectrum. Finally, we state which
subsample each source belongs to, and we indicate the sources that were not used in the creation of the composite templates.

\clearpage
\centering


\section*{Supplementary material}

TABLE 1
MIPS AND *Herschel* PHOTOMETRY

| ID | RA | Dec | Subsample | AGN (%) [a] | $z$ [b] | $S_{24}$ (μJy) | $S_{70}$ (mJy) | $S_{100}$ (mJy) | $S_{160}$ (mJy) | $S_{250}$ (mJy) [c] | $S_{350}$ (mJy) [c] | $S_{500}$ (mJy) [c] |
|---|---|---|---|---|---|---|---|---|---|---|---|---|
| GN_IRS1 | 12:36:45.83 | +62:07:54.0 | $z \sim 1$ SF | 13 | 1.43 | 374.0 ± 7.9 | 5.62 ± 0.71 | 14.66 ± 0.41 | 21.53 ± 0.64 | 20.91 ± 1.77 | 12.40 ± 1.91 | *0.40 ± 2.00* |
| GN_IRS2 | 12:37:02.74 | +62:14:01.0 | $z \sim 1$ SF | 0 | 1.24 | 349.0 ± 8.8 | 2.41 ± 0.70 | 7.89 ± 0.29 | 16.89 ± 0.77 | 25.01 ± 1.80 | 18.15 ± 2.35 | 11.77 ± 2.74 |
| GN_IRS3 | 12:37:10.62 | +62:22:34.0 | $z \sim 2$ SF | 0 | 1.52 | 386.0 ± 4.9 | 4.88 ± 0.68 | 8.02 ± 0.48 | 15.52 ± 0.66 | 24.46 ± 2.11 | 20.41 ± 2.53 | *7.00 ± 2.00* |
| GN_IRS4 | 12:36:53.37 | +62:11:39.0 | $z \sim 1$ SF | 17 | 1.27 | 336.0 ± 7.5 | 7.01 ± 0.73 | 16.14 ± 0.32 | 24.21 ± 0.64 | 20.39 ± 1.25 | 12.24 ± 3.45 | *9.00 ± 2.00* |
| GN_IRS5 | 12:36:20.94 | +62:07:14.0 | $z \sim 1$ SF | 0 | 1.15 | 451.0 ± 8.8 | 5.34 ± 0.68 | 13.23 ± 0.42 | 24.26 ± 1.87 | 33.91 ± 4.90 | 23.25 ± 9.30 | *12.60 ± 2.00* |
| GN_IRS6 | 12:36:33.68 | +62:10:05.0 | $z \sim 1$ SF | 0 | 1.02 | 581.0 ± 9.0 | 2.74 ± 0.65 | 8.52 ± 0.54 | 16.54 ± 0.70 | 20.24 ± 1.87 | 11.38 ± 1.21 | ... |
| GN_IRS7[d] | 12:37:11.98 | +62:13:25.7 | $z \sim 2$ SF | 0 | 1.99 | 219.0 ± 6.6 | ... | 3.65 ± 0.40 | 6.48 ± 0.65 | *0.50 ± 2.00* | ... | *5.60 ± 2.00* |
| GN_IRS8[d] | 12:37:21.87 | +62:10:35.3 | Sil AGN | 61 | 0.98 | 255.0 ± 5.5 | ... | 2.00 ± 0.38 | 1.77 ± 0.65 | *25.00 ± 4.29* | *5.10 ± 2.00* | *33.00 ± 5.57* |
| GN_IRS9 | 12:35:49.44 | +62:15:36.8 | $z \sim 2$ SF | 0 | 2.20 | ... | ... | ... | ... | *32.00 ± 5.02* | *5.02* | 23.78 ± 2.67 |
| GN_IRS10 | 12:37:07.21 | +62:14:08.1 | $z \sim 2$ SF | 0 | 2.49 | 253.0 ± 8.2 | ... | 1.20 ± 0.36 | 6.82 ± 0.95 | 23.95 ± 2.37 | 26.38 ± 2.70 | 17.02 ± 3.57 |
| GN_IRS11 | 12:36:21.27 | +62:17:08.0 | $z \sim 2$ SF | 23 | 2.00 | 370.0 ± 11.0 | ... | 3.59 ± 0.38 | 12.67 ± 1.07 | 25.00 ± 1.83 | 25.86 ± 1.72 | *17.02 ± 3.57* |
| GN_IRS12 | 12:36:00.16 | +62:10:47.0 | Sil AGN | 79 | 1.99 | 1240.0 ± 10.2 | 4.03 ± 0.56 | 14.17 ± 0.46 | 13.65 ± 1.06 | 17.06 ± 2.31 | 16.27 ± 1.59 | 14.70 ± 6.97 |
| GN_IRS13 | 12:37:34.52 | +62:17:23.0 | $z \sim 1$ SF | 2 | 0.64 | 741.0 ± 7.1 | 5.53 ± 0.70 | 16.46 ± 0.27 | 33.35 ± 0.73 | 36.89 ± 0.94 | 22.73 ± 1.60 | 5.78 ± 1.81 |
| GN_IRS14 | 12:36:22.48 | +62:15:44.0 | $z \sim 1$ SF | 3 | 0.64 | 725.0 ± 6.8 | 11.10 ± 0.72 | 21.02 ± 0.40 | 27.01 ± 0.82 | 22.19 ± 1.56 | 11.27 ± 1.39 | 5.59 ± 2.17 |
| GN_IRS15 | 12:37:11.37 | +62:13:31.0 | $z \sim 2$ SF | 0 | 2.00 | 537.0 ± 9.3 | ... | 7.20 ± 0.39 | 26.47 ± 0.62 | 42.85 ± 1.80 | 41.69 ± 2.15 | 31.98 ± 3.07 |
| GN_IRS16 | 12:36:37.02 | +62:08:52.0 | $z \sim 2$ SF | 10 | 2.02 | 371.0 ± 7.7 | ... | 1.90 ± 0.29 | 5.90 ± 0.74 | 15.31 ± 2.94 | 15.17 ± 0.83 | *18.00 ± 2.00* |
| GN_IRS17 | 12:36:16.11 | +62:15:13.0 | $z \sim 2$ SF | 47 | 2.57 | 326.0 ± 8.0 | ... | 5.53 ± 0.39 | 13.59 ± 0.95 | 28.71 ± 1.65 | 27.40 ± 1.66 | 15.02 ± 3.90 |
| GN_IRS18 | 12:37:16.59 | +62:16:43.0 | $z \sim 2$ SF | 29 | 1.80 | 508.0 ± 6.3 | ... | 4.72 ± 0.25 | 13.90 ± 0.82 | 22.99 ± 1.52 | 17.28 ± 1.70 | 9.44 ± 3.42 |
| GN_IRS19 | 12:35:55.13 | +62:09:01.0 | Sil AGN | 63 | 1.88 | 373.0 ± 8.7 | ... | 3.15 ± 0.54 | 5.03 ± 1.29 | *12.00 ± 2.00* | *6.00 ± 2.00* | ... |
| GN_IRS20 | 12:36:03.25 | +62:11:10.0 | $z \sim 1$ SF | 1 | 0.64 | 1250.0 ± 10.1 | 11.10 ± 0.72 | 30.42 ± 0.75 | 34.80 ± 0.79 | 23.95 ± 2.42 | 10.22 ± 1.69 | *1.50 ± 5.57* |
| GN_IRS21 | 12:36:18.33 | +62:15:50.0 | $z \sim 2$ SF | 36 | 2.00 | 311.0 ± 5.3 | ... | 4.97 ± 0.34 | 18.89 ± 0.93 | 44.09 ± 1.94 | 45.34 ± 1.27 | 22.47 ± 3.01 |
| GN_IRS22[d] | 12:37:04.34 | +62:14:46.0 | Sil AGN | 95 | 2.21 | 380.0 ± 4.8 | 1.59 ± 0.56 | 4.10 ± 0.40 | 2.98 ± 0.90 | 4.56 ± 1.79 | *18.00 ± 2.00* | 11.25 ± 4.61 |
| GN_IRS23 | 12:36:19.13 | +62:14:04.0 | $z \sim 2$ SF | 13 | 2.21 | 219.0 ± 6.9 | ... | 1.27 ± 0.28 | 2.64 ± 0.71 | 11.33 ± 1.96 | 11.77 ± 3.21 | 9.50 ± 2.77 |
| GN_IRS24[e] | 12:36:53.22 | +62:11:16.0 | Sil AGN | 54 | 0.93 | 363.0 ± 7.1 | ... | 2.76 ± 0.33 | 8.74 ± 0.72 | 13.77 ± 1.29 | 35.00 ± 2.34 | 28.82 ± 2.93 |
| GN_IRS25 | 12:37:01.59 | +62:11:46.0 | $z \sim 2$ SF | 7 | 1.72 | 739.0 ± 10.7 | 1.31 ± 0.65 | 6.98 ± 0.46 | 19.44 ± 0.91 | 38.38 ± 1.46 | 35.60 ± 2.34 | 19.63 ± 2.33 |
| GN_IRS26 | 12:36:34.51 | +62:12:40.0 | $z \sim 2$ SF | 4 | 1.22 | 444.0 ± 5.5 | 13.20 ± 0.66 | 34.20 ± 0.46 | 63.61 ± 1.71 | 62.50 ± 1.62 | 45.60 ± 2.63 | ... |
| GN_IRS27 | 12:36:33.22 | +62:08:34.0 | Sil AGN | 26 | 0.79 | 859.0 ± 10.8 | 14.20 ± 0.71 | 28.37 ± 0.46 | 41.40 ± 0.86 | 30.97 ± 1.68 | 17.22 ± 2.63 | *5.70 ± 2.00* |
| GN_IRS28 | 12:36:56.47 | +62:19:37.0 | Sil AGN | 54 | 0.94 | 779.0 ± 7.9 | 4.46 ± 0.62 | 10.80 ± 0.32 | 21.01 ± 0.61 | 20.64 ± 3.73 | 11.82 ± 0.97 | *0.40 ± 2.00* |
| GN_IRS29 | 12:36:33.22 | +62:20:26.0 | Sil AGN | 74 | 2.29 | 716.0 ± 7.4 | 2.21 ± 0.64 | 7.55 ± 0.26 | 12.88 ± 0.62 | 10.91 ± 2.21 | 8.54 ± 2.61 | *4.50 ± 2.00* |
| GN_IRS30 | 12:36:56.49 | +62:20:26.0 | Sil AGN | 66 | 1.76 | 931.0 ± 8.6 | 3.44 ± 0.66 | 10.19 ± 0.40 | 16.89 ± 0.75 | 15.63 ± 1.11 | 24.37 ± 1.52 | 22.69 ± 2.40 |
| GN_IRS31[e] | 12:36:22.66 | +62:16:29.0 | $z \sim 2$ SF | 0 | 1.79 | 417.0 ± 8.2 | ... | 2.85 ± 0.35 | 10.34 ± 1.12 | 25.23 ± 1.79 | 17.28 ± 1.70 | 10.83 ± 4.57 |
| GN_IRS32 | 12:37:18.27 | +62:22:58.0 | $z \sim 2$ SF | 49 | 1.89 | 483.0 ± 5.7 | ... | 2.32 ± 0.29 | 5.78 ± 1.09 | 12.57 ± 1.25 | 12.37 ± 2.06 | *9.00 ± 2.00* |
| GN_IRS33[e] | 12:36:42.21 | +62:15:45.9 | $z \sim 1$ SF | 6 | 0.86 | 866.0 ± 10.3 | 4.56 ± 0.63 | 10.99 ± 0.32 | 21.40 ± 0.71 | 23.36 ± 2.06 | 9.66 ± 1.64 | ... |
| GN_IRS34[d] | 12:37:39.21 | +62:20:59.6 | $z \sim 1$ SF | 0 | 0.84 | 760.0 ± 8.2 | 2.60 ± 0.62 | 7.53 ± 0.35 | 10.20 ± 1.22 | 14.36 ± 2.74 | *10.00 ± 2.00* | ... |
| GN_IRS35[d] | 12:36:32.58 | +62:08:00.1 | Feat AGN | 95 | 2.00 | 820.0 ± 9.6 | 5.86 ± 0.74 | 12.50 ± 0.24 | 24.14 ± 0.85 | 28.07 ± 1.40 | *32.00 ± 2.00* | 25.06 ± 3.13 |
| GN_IRS36 | 12:36:35.59 | +62:14:24.5 | Feat AGN | 93 | 2.02 | 1520.0 ± 14.4 | 2.37 ± 0.69 | 3.42 ± 0.36 | 3.74 ± 0.73 | 3.30 ± 1.22 | ... | *1.30 ± 2.00* |
| GN_IRS37 | 12:37:16.66 | +62:17:33.5 | Feat AGN | 100 | 1.15 | 1240.0 ± 15.5 | ... | 7.58 ± 0.25 | 19.89 ± 0.72 | 31.61 ± 2.05 | 24.43 ± 1.56 | 14.73 ± 2.31 |
| GN_IRS38 | 12:36:29.10 | +62:10:46.3 | $z \sim 1$ SF | 0 | 1.01 | 730.0 ± 12.7 | 7.60 ± 0.79 | 10.97 ± 0.29 | 7.28 ± 0.93 | 5.57 ± 1.43 | *4.00 ± 2.00* | *3.10 ± 2.00* |
| GN_IRS39 | 12:36:08.11 | +62:10:36.4 | Feat AGN | 100 | 0.68 | 2300.0 ± 13.5 | 6.23 ± 0.79 | 9.45 ± 0.37 | 6.47 ± 0.67 | ... | ... | ... |
| GN_IRS40 | 12:36:49.65 | +62:07:38.6 | Feat AGN | 100 | 1.60 | 1440.0 ± 12.6 | 8.93 ± 0.72 | 15.29 ± 0.43 | 14.01 ± 1.02 | *9.00 ± 2.00* | *7.00 ± 2.00* | ... |
| GN_IRS41 | 12:36:29.98 | +62:05:42.8 | Feat AGN | 80 | 0.48 | 2480.0 ± 26.0 | 5.38 ± 0.68 | 17.81 ± 0.41 | 38.74 ± 1.38 | 33.69 ± 2.11 | 24.16 ± 1.66 | 15.54 ± 3.13 |
| GN_IRS42 | 12:36:46.72 | +62:08:33.9 | $z \sim 1$ SF | 0 | 0.97 | 991.0 ± 8.8 | ... | 1.17 ± 0.35 | ... | ... | ... | ... |
| GN_IRS43 | 12:37:04.02 | +62:11:57.6 | ... | 84 | 3.40 | 105.0 ± 9.5 | ... | ... | ... | ... | ... | ... |
| GN_IRS44 | 12:37:37.06 | +62:12:05.0 | $z \sim 1$ SF | 10 | 0.40 | 763.0 ± 8.0 | 3.06 ± 0.73 | 15.97 ± 0.45 | 26.56 ± 0.86 | 26.63 ± 1.92 | 12.26 ± 3.54 | *23.00 ± 2.00* |
| GN_IRS45[d] | 12:38:21.76 | +62:17:06.0 | $z \sim 2$ SF | 43 | 1.62 | 1010.0 ± 12.8 | ... | 13.15 ± 0.88 | 16.23 ± 1.82 | *27.00 ± 2.00* | *25.00 ± 2.00* | ... |
| GN_IRS46 | 12:35:48.10 | +62:10:35.0 | $z \sim 1$ SF | 0 | 0.64 | 568.0 ± 12.4 | 4.33 ± 0.60 | 4.56 ± 0.41 | 11.14 ± 0.94 | 14.57 ± 2.34 | 34.49 ± 4.72 | 29.29 ± 4.06 |
| GN_IRS47[e] | 12:35:53.81 | +62:13:38.0 | $z \sim 1$ SF | 0 | 0.88 | 553.0 ± 13.0 | ... | 8.17 ± 0.49 | 15.50 ± 0.83 | 35.60 ± 2.00 | 23.00 ± 2.00 | *13.00 ± 2.00* |
| GN_IRS48 | 12:36:51.58 | +62:06:44.0 | $z \sim 1$ SF | 0 | 0.84 | 582.0 ± 11.4 | 1.88 ± 0.66 | ... | ... | *20.00 ± 2.00* | *4.00 ± 2.00* | ... |
| GN_IRS49[d] | 12:37:05.49 | +62:21:24.0 | $z \sim 1$ SF | 0 | 0.95 | 513.0 ± 7.5 | ... | 3.93 ± 0.42 | 10.33 ± 0.89 | 15.43 ± 4.45 | 12.37 ± 4.19 | 11.06 ± 3.69 |
| GN_IRS50 | 12:37:25.48 | +62:18:50.0 | $z \sim 1$ SF | 0 | 0.84 | 484.0 ± 6.2 | 1.22 ± 0.63 | 6.18 ± 0.38 | 12.25 ± 0.55 | 21.16 ± 2.24 | 22.57 ± 2.15 | *15.00 ± 2.00* |
| GN_IRS51[e] | 12:36:24.46 | +62:08:36.0 | $z \sim 1$ SF | 0 | 0.95 | 695.0 ± 11.5 | 2.40 ± 0.70 | 4.48 ± 0.35 | 11.88 ± 0.98 | 8.93 ± 1.36 | 6.82 ± 0.95 | *8.00 ± 2.00* |
| GN_IRS52[e] | 12:37:27.83 | +62:14:37.0 | $z \sim 1$ SF | 0 | 0.94 | 338.0 ± 6.1 | ... | 2.17 ± 0.36 | 6.53 ± 0.49 | ... | ... | *8.60 ± 2.00* |
| GN_IRS53[d] | 12:36:09.65 | +62:08:45.0 | $z \sim 1$ SF | 1 | 0.91 | 370.0 ± 6.4 | ... | 2.27 ± 0.41 | 7.80 ± 0.84 | 14.96 ± 1.50 | 8.07 ± 3.39 | *6.00 ± 2.00* |
| GN_IRS54 | 12:37:35.29 | +62:19:18.0 | $z \sim 1$ SF | 34 | 0.84 | 496.0 ± 6.6 | ... | 2.71 ± 0.31 | 5.90 ± 0.88 | 7.22 ± 2.15 | ... | ... |

TABLE 1 — Continued

| ID | RA | Dec | Subsample | AGN (%) [a] | $z$ [b] | $S_{24}$ (μJy) | $S_{70}$ (mJy) | $S_{100}$ (mJy) | $S_{160}$ (mJy) | $S_{250}$ (mJy) [c] | $S_{350}$ (mJy) [c] | $S_{500}$ (mJy) [c] |
|---|---|---|---|---|---|---|---|---|---|---|---|---|
| GN_IRS55 | 12:36:46.73 | +62:14:46.3 | Sil AGN | 66 | 1.99 | 422.0 ± 29.1 | ... | 4.16 ± 0.42 | 7.69 ± 0.98 | 7.01 ± 2.06 | 10.62 ± 2.12 | 9.71 ± 3.76 |
| GN_IRS56 | 12:36:17.78 | +62:14:08.0 | $z \sim 2$ SF | 14 | 1.67 | 429.0 ± 7.4 | ... | 3.17 ± 0.33 | 8.62 ± 0.61 | 14.05 ± 1.76 | ... | ... |
| GN_IRS57[f] | 12:36:40.74 | +62:10:10.6 | Feat AGN | 90 | 1.96 | ... | ... | ... | ... | ... | ... | ... |
| GN_IRS58[d] | 12:36:44.83 | +62:17:16.0 | $z \sim 2$ SF | 30 | 2.04 | 167.0 ± 5.0 | ... | ... | 3.33 ± 0.72 | ... | ... | ... |
| GN_IRS59[d] | 12:36:44.13 | +62:14:50.7 | Sil AGN | 75 | 1.77 | 141.0 ± 5.9 | ... | 2.40 ± 0.36 | 4.85 ± 0.73 | ... | ... | ... |
| GN_IRS60 | 12:36:49.72 | +62:13:12.9 | $z \sim 1$ SF | 0 | 0.47 | 337.0 ± 10.9 | 6.41 ± 0.67 | 13.39 ± 0.30 | 24.15 ± 0.62 | 20.60 ± 2.02 | 18.00 ± 2.00 | 10.00 ± 2.00 |
| GN_IRS61 | 12:36:17.33 | +62:15:29.6 | $z \sim 1$ SF | 0 | 0.85 | 502.0 ± 7.7 | ... | 6.06 ± 0.31 | 12.58 ± 1.07 | 9.55 ± 1.57 | ... | ... |
| GN_IRS62 | 12:36:29.54 | +62:06:46.5 | $z \sim 1$ SF | 13 | 0.80 | 372.0 ± 6.2 | ... | 6.87 ± 0.40 | 6.83 ± 1.01 | 11.71 ± 3.94 | 7.12 ± 1.40 | 4.00 ± 2.00 |
| GN_IRS63 | 12:36:26.58 | +62:08:35.7 | $z \sim 1$ SF | 9 | 1.47 | 139.0 ± 6.4 | ... | 2.97 ± 0.33 | 9.25 ± 0.96 | 14.46 ± 2.22 | 7.59 ± 1.24 | ... |
| GN_IRS64 | 12:37:11.90 | +62:22:12.1 | $z \sim 1$ SF | 66 | 4.05 | 65.5 ± 4.4 | ... | ... | 4.41 ± 0.92 | 12.86 ± 1.75 | 16.70 ± 2.61 | 29.66 ± 5.81 |
| GN_IRS65 | 12:36:45.54 | +62:08:25.1 | ... | 59 | 0.47 | 236.0 ± 8.4 | 2.94 ± 0.70 | 5.75 ± 0.42 | 9.76 ± 1.32 | 14.02 ± 2.00 | 8.40 ± 1.65 | ... |
| GN_IRS66[e] | 12:37:05.85 | +62:21:29.8 | $z \sim 1$ SF | 3 | 0.95 | 364.0 ± 5.5 | ... | 5.36 ± 0.38 | 11.02 ± 0.95 | ... | ... | ... |
| GN_IRS67[d] | 12:36:19.14 | +62:13:01.8 | $z \sim 1$ SF | 0 | 1.23 | 220.0 ± 8.2 | ... | 5.24 ± 0.30 | 11.46 ± 1.48 | ... | ... | ... |
| GN_IRS68[e] | 12:36:19.50 | +62:12:52.6 | $z \sim 1$ SF | 0 | 0.47 | 971.0 ± 10.9 | 16.30 ± 0.77 | 28.89 ± 0.30 | 31.99 ± 1.42 | 31.61 ± 1.05 | 14.70 ± 1.98 | 8.60 ± 2.00 |
| GN_IRS69[e] | 12:35:47.70 | +62:10:58.2 | $z \sim 1$ SF | 13 | 0.45 | 698.0 ± 10.0 | 13.20 ± 0.66 | 24.31 ± 0.62 | 34.41 ± 1.36 | 24.80 ± 2.93 | 11.60 ± 2.00 | 8.00 ± 2.00 |
| GN_IRS70[e] | 12:35:41.39 | +62:12:17.8 | $z \sim 1$ SF | 0 | 0.46 | 419.0 ± 8.9 | 2.59 ± 0.62 | 9.08 ± 0.38 | 11.80 ± 1.11 | 23.00 ± 2.00 | 18.00 ± 2.00 | 13.94 ± 2.92 |
| GS_IRS1 | 12:35:44.00 | 03:32:44.00 | $z \sim 2$ SF | 38 | 2.69 | 104.0 ± 6.2 | ... | 1.88 ± 0.35 | 6.11 ± 1.11 | 17.32 ± 2.98 | 21.86 ± 2.98 | 13.94 ± 2.92 |
| GS_IRS2 | 03:32:34.85 | -27:46:35.0 | $z \sim 1$ SF | 0 | 1.10 | 115.0 ± 6.9 | ... | 2.25 ± 0.28 | 4.59 ± 1.24 | 6.50 ± 2.00 | 5.10 ± 2.00 | 2.50 ± 2.00 |
| GS_IRS3 | 03:32:08.66 | -27:47:34.4 | Feat AGN | 100 | 0.55 | 3560.0 ± 36.0 | 2.69 ± 0.50 | 3.61 ± 0.44 | 3.75 ± 1.63 | 3.62 ± 2.28 | 4.54 ± 2.85 | 0.98 ± 6.48 |
| GS_IRS4[d] | 03:33:01.45 | -28:02:51.3 | Sil AGN | 99 | 1.30 | ... | ... | ... | ... | 12.28 ± 2.40 | 7.93 ± 2.42 | ... |
| GS_IRS5 | 03:32:21.07 | -27:35:30.0 | Sil AGN | 68 | 0.70 | ... | 3.99 ± 0.89 | ... | ... | 73.24 ± 2.33 | 49.37 ± 2.30 | 13.07 ± 2.58 |
| GS_IRS6 | 03:32:49.19 | -27:40:50.6 | Feat AGN | 100 | 0.55 | 2270.0 ± 23.0 | 5.22 ± 0.65 | 4.70 ± 0.30 | ... | 11.73 ± 2.24 | 16.14 ± 4.25 | 2.20 ± 2.24 |
| GS_IRS7 | 03:32:26.49 | -27:40:35.7 | Sil AGN | 89 | 1.03 | 2250.0 ± 22.8 | 4.85 ± 0.74 | ... | ... | 25.58 ± 3.06 | 32.77 ± 2.37 | 7.70 ± 3.02 |
| GS_IRS8 | 03:32:11.65 | -27:37:25.9 | Sil AGN | 82 | 1.58 | 1830.0 ± 18.5 | ... | ... | ... | 40.77 ± 2.26 | 32.77 ± 2.37 | 14.50 ± 2.00 |
| GS_IRS9 | 03:32:29.86 | -27:51:05.8 | ... | 80 | 3.71 | 84.6 ± 4.7 | 2.94 ± 0.82 | 0.58 ± 0.15 | ... | 2.51 ± 2.62 | ... | ... |
| GS_IRS10 | 03:31:58.27 | -27:50:41.0 | Feat AGN | 98 | 1.15 | 1240.0 ± 12.1 | 2.91 ± 0.81 | ... | ... | 16.80 ± 2.27 | 13.19 ± 2.31 | 11.48 ± 2.47 |
| GS_IRS11[e] | 03:31:46.61 | -27:52:12.0 | Feat AGN | 100 | 0.68 | 864.0 ± 7.9 | 3.70 ± 0.69 | 4.41 ± 0.59 | 5.32 ± 0.88 | 10.47 ± 2.29 | 10.36 ± 2.35 | 2.50 ± 2.00 |
| GS_IRS12 | 03:32:37.76 | -28:01:09.0 | Feat AGN | 100 | 1.60 | 1070.0 ± 10.4 | ... | ... | ... | 7.35 ± 3.06 | 5.47 ± 2.60 | 9.61 ± 2.52 |
| GS_IRS13 | 03:32:10.48 | -27:46:44.0 | Sil AGN | 76 | 1.62 | 752.0 ± 8.1 | ... | ... | ... | 10.28 ± 2.35 | 15.00 ± 2.32 | 7.00 ± 2.00 |
| GS_IRS14 | 03:32:14.47 | -27:49:26.0 | Sil AGN | 50 | 1.87 | 183.0 ± 6.5 | 4.02 ± 0.54 | 6.37 ± 0.29 | 5.48 ± 1.80 | 9.94 ± 4.50 | 11.57 ± 4.95 | 0.78 ± 2.96 |
| GS_IRS15 | 03:32:40.74 | -27:46:40.0 | $z \sim 2$ SF | 23 | 2.11 | 192.0 ± 6.8 | ... | 1.58 ± 0.44 | ... | 8.63 ± 2.24 | 6.57 ± 2.32 | 6.70 ± 2.00 |
| GS_IRS16 | 03:33:15.66 | -27:46:07.0 | Sil AGN | 0 | 0.83 | 285.0 ± 6.2 | 3.62 ± 0.81 | 1.63 ± 0.19 | 2.83 ± 0.78 | 7.50 ± 2.00 | 9.40 ± 2.00 | ... |
| GS_IRS17[d] | 03:31:17.46 | -27:47:55.0 | $z \sim 1$ SF | 0 | 1.01 | ... | ... | ... | ... | 7.47 ± 2.29 | 7.35 ± 2.32 | 2.10 ± 2.00 |
| GS_IRS18 | 03:31:16.37 | -27:45:33.0 | $z \sim 1$ SF | 0 | 1.04 | ... | ... | ... | ... | 6.50 ± 2.00 | 5.35 ± 2.42 | 3.00 ± 2.00 |
| GS_IRS19 | 03:32:17.63 | -27:45:51.0 | $z \sim 2$ SF | 10 | 1.83 | 167.0 ± 5.9 | ... | ... | 2.06 ± 0.71 | 4.05 ± 2.62 | 4.23 ± 2.47 | ... |
| GS_IRS20 | 03:32:47.58 | -27:44:52.0 | $z \sim 2$ SF | 24 | 1.91 | 275.0 ± 6.0 | 3.17 ± 0.59 | 8.32 ± 0.56 | 20.15 ± 1.60 | 24.78 ± 2.30 | 23.82 ± 3.08 | 12.49 ± 5.08 |
| GS_IRS21 | 03:32:42.27 | -27:47:46.0 | $z \sim 1$ SF | 0 | 0.99 | 247.0 ± 8.8 | ... | 2.71 ± 0.23 | 4.89 ± 0.98 | 6.70 ± 3.12 | 4.34 ± 3.00 | ... |
| GS_IRS22 | 03:32:20.05 | -27:44:47.0 | Sil AGN | 84 | 1.90 | 338.0 ± 5.8 | 4.02 ± 0.54 | 6.37 ± 0.29 | 7.52 ± 1.08 | 6.55 ± 2.00 | 5.40 ± 2.00 | 2.15 ± 2.00 |
| GS_IRS23 | 03:32:17.23 | -27:50:37.0 | $z \sim 2$ SF | 0 | 1.96 | 144.0 ± 4.9 | ... | 1.58 ± 0.44 | 5.16 ± 1.94 | 6.55 ± 2.00 | 19.93 ± 3.33 | 13.04 ± 6.62 |
| GS_IRS24 | 03:33:04.38 | -27:48:55.0 | Sil AGN | 100 | 0.96 | 287.0 ± 6.2 | 3.02 ± 0.84 | 5.09 ± 0.22 | 11.10 ± 0.93 | 17.11 ± 2.23 | 16.69 ± 2.29 | 10.72 ± 2.60 |
| GS_IRS25 | 03:32:18.70 | -27:49:19.0 | $z \sim 1$ SF | 0 | 1.04 | 221.0 ± 7.9 | 2.15 ± 0.52 | 2.32 ± 0.24 | 6.88 ± 1.24 | 15.59 ± 2.25 | 10.67 ± 2.63 | 3.99 ± 6.20 |
| GS_IRS26 | 03:32:20.70 | -27:44:53.0 | $z \sim 1$ SF | 0 | 0.97 | 288.0 ± 6.2 | ... | ... | 6.88 ± 1.24 | 14.02 ± 2.46 | 10.67 ± 2.63 | 1.50 ± 2.00 |
| GS_IRS27 | 03:32:41.88 | -27:52:44.0 | $z \sim 2$ SF | 37 | 1.75 | 278.0 ± 8.7 | 3.47 ± 0.65 | ... | 5.34 ± 1.25 | 10.84 ± 2.94 | 10.05 ± 2.80 | 3.90 ± 2.00 |
| GS_IRS28 | 03:31:35.26 | -27:49:58.0 | $z \sim 1$ SF | 0 | 0.96 | 232.0 ± 8.3 | ... | ... | ... | 10.60 ± 2.00 | 7.90 ± 2.00 | 3.40 ± 2.00 |
| GS_IRS29[f] | 03:32:22.53 | -27:45:38.0 | $z \sim 2$ SF | 42 | 2.08 | 147.0 ± 5.0 | ... | 1.08 ± 0.30 | 2.45 ± 1.41 | 8.80 ± 2.00 | 7.15 ± 2.00 | ... |
| GS_IRS30[d] | 03:32:43.78 | -27:52:31.0 | $z \sim 1$ SF | 0 | 1.62 | 139.0 ± 4.7 | ... | ... | ... | 5.67 ± 2.84 | 6.23 ± 2.73 | 6.34 ± 3.00 |
| GS_IRS31 | 03:32:06.00 | -27:45:15.0 | $z \sim 1$ SF | 0 | 1.07 | 336.0 ± 6.4 | 2.02 ± 0.83 | 5.11 ± 0.66 | 10.74 ± 1.79 | 10.28 ± 2.66 | 6.51 ± 2.68 | 6.50 ± 3.46 |
| GS_IRS32[d] | 03:32:30.19 | -27:57:01.0 | $z \sim 1$ SF | 36 | 0.78 | 191.0 ± 7.2 | ... | 2.96 ± 0.33 | 4.56 ± 1.84 | 8.85 ± 3.12 | 9.88 ± 9.30 | 2.74 ± 12.52 |
| GS_IRS33 | 03:32:23.43 | -27:42:55.0 | Sil AGN | 100 | 2.14 | 220.0 ± 7.8 | ... | ... | 2.05 ± 0.88 | 3.96 ± 2.80 | 10.64 ± 2.44 | 9.40 ± 3.38 |
| GS_IRS34 | 03:32:39.00 | -27:44:20.0 | Sil AGN | 53 | 1.93 | 180.0 ± 6.4 | ... | 4.79 ± 0.37 | 11.91 ± 1.10 | 6.01 ± 2.48 | 5.03 ± 4.38 | 8.56 ± 2.86 |
| GS_IRS35 | 03:32:43.00 | -27:46:50.0 | $z \sim 1$ SF | 0 | 1.04 | 201.0 ± 7.2 | ... | 4.79 ± 0.25 | 7.66 ± 1.30 | 15.42 ± 3.32 | 12.12 ± 3.08 | 10.75 ± 2.86 |
| GS_IRS36 | 03:32:45.90 | -27:52:20.0 | $z \sim 1$ SF | 0 | 0.97 | 200.0 ± 7.1 | ... | ... | ... | 17.73 ± 3.66 | 12.32 ± 2.93 | 8.47 ± 3.22 |
| GS_IRS37 | 03:32:16.86 | -27:54:18.0 | $z \sim 1$ SF | 26 | 0.96 | 259.0 ± 5.6 | 2.78 ± 0.78 | 2.57 ± 0.50 | ... | 7.40 ± 2.00 | 5.68 ± 3.20 | 3.30 ± 2.00 |
| GS_IRS38 | 03:32:25.68 | -27:43:05.0 | Sil AGN | 88 | 2.30 | 516.0 ± 6.8 | ... | ... | ... | 7.15 ± 2.29 | 10.85 ± 2.32 | 9.47 ± 3.38 |
| GS_IRS39[d] | 03:31:48.18 | -27:45:35.0 | $z \sim 2$ SF | 0 | 1.72 | ... | ... | ... | ... | 13.73 ± 2.22 | 18.95 ± 2.29 | 12.00 ± 2.00 |
| GS_IRS40[d] | 03:33:27.80 | -28:02:51.0 | $z \sim 1$ SF | 0 | 0.94 | 367.0 ± 6.4 | ... | ... | ... | 7.30 ± 2.00 | 6.40 ± 2.00 | 3.40 ± 2.00 |

TABLE 1 — Continued

| ID | RA | Dec | Subsample | AGN (%)[a] | $z$[b] | $S_{24}$ (μJy) | $S_{70}$ (mJy) | $S_{100}$ (mJy) | $S_{160}$ (mJy) | $S_{250}$ (mJy)[c] | $S_{350}$ (mJy)[c] | $S_{500}$ (mJy)[c] |
|---|---|---|---|---|---|---|---|---|---|---|---|---|
| GS_IRS41 | 03:31:51.50 | -27:41:57.0 | $z \sim 1$ SF | 0 | 0.86 | 291.0 ± 8.7 | 2.38 ± 0.98 | ... | ... | 11.29 ± 2.29 | 13.00 ± 2.00 | 7.40 ± 2.00 |
| GS_IRS42 | 03:32:25.98 | -27:47:51.0 | Sil AGN | 66 | 1.88 | 144.0 ± 4.9 | ... | 1.60 ± 0.24 | 3.34 ± 0.80 | 4.55 ± 2.40 | 15.33 ± 2.30 | ... |
| GS_IRS43 | 03:31:48.93 | -27:39:45.0 | $z \sim 1$ SF | 0 | 0.82 | 446.0 ± 9.3 | 2.37 ± 0.97 | ... | ... | 20.24 ± 2.24 | ... | ... |
| GS_IRS44 | 03:32:25.41 | -27:46:17.0 | $z \sim 1$ SF | 0 | 0.90 | 349.0 ± 6.0 | 2.95 ± 0.55 | 5.19 ± 0.31 | 8.93 ± 1.11 | 10.47 ± 2.24 | 6.82 ± 3.23 | 3.51 ± 2.88 |
| GS_IRS45 | 03:32:17.45 | -27:50:03.0 | $z \sim 2$ SF | 0 | 1.62 | 308.0 ± 6.7 | ... | 7.07 ± 0.31 | 19.04 ± 1.12 | 33.03 ± 2.40 | 31.98 ± 2.53 | 10.98 ± 6.44 |
| GS_IRS46[d] | 03:32:42.71 | -27:39:27.0 | $z \sim 2$ SF | 0 | 1.85 | 378.0 ± 6.6 | ... | ... | ... | *13.05 ± 2.80* | *15.11 ± 4.70* | *13.00 ± 2.00* |
| GS_IRS47 | 03:32:32.88 | -27:41:24.0 | $z \sim 1$ SF | 0 | 1.04 | 212.0 ± 7.6 | ... | 3.21 ± 0.98 | ... | 17.05 ± 2.68 | 13.49 ± 3.00 | 10.10 ± 3.32 |
| GS_IRS48[d] | 03:31:30.00 | -27:35:22.0 | $z \sim 1$ SF | 0 | 0.95 | ... | ... | ... | ... | 9.04 ± 2.65 | 10.07 ± 2.79 | ... |
| GS_IRS49[d] | 03:32:48.51 | -27:54:16.0 | $z \sim 1$ SF | 0 | 0.83 | 184.0 ± 6.5 | ... | ... | ... | 17.54 ± 2.28 | 20.07 ± 2.47 | 21.80 ± 3.12 |
| GS_IRS50 | 03:32:31.52 | -27:48:53.0 | $z \sim 2$ SF | 5 | 1.90 | 227.0 ± 8.1 | ... | ... | ... | 5.83 ± 2.84 | 6.06 ± 2.45 | 4.03 ± 3.44 |
| GS_IRS51[d] | 03:32:23.71 | -27:44:11.0 | $z \sim 2$ SF | 11 | 1.77 | 241.0 ± 8.6 | ... | 1.12 ± 0.13 | 3.62 ± 1.05 | 0.74 ± 2.72 | 5.71 ± 3.10 | 1.70 ± 3.16 |
| GS_IRS52 | 03:32:12.52 | -27:43:06.0 | $z \sim 2$ SF | 1 | 1.79 | 240.0 ± 8.6 | ... | 1.15 ± 0.40 | 3.51 ± 1.12 | 10.06 ± 3.26 | 9.05 ± 4.40 | 6.86 ± 3.66 |
| GS_IRS53 | 03:32:12.13 | -27:42:49.0 | $z \sim 2$ SF | 0 | 2.45 | 152.0 ± 5.1 | ... | ... | ... | 9.26 ± 3.52 | 12.01 ± 4.43 | 7.15 ± 3.54 |
| GS_IRS54 | 03:32:13.85 | -27:43:12.0 | $z \sim 2$ SF | 40 | 1.89 | 249.0 ± 8.9 | ... | 3.77 ± 0.73 | 3.96 ± 2.85 | 15.52 ± 3.20 | 15.73 ± 4.25 | 15.42 ± 3.72 |
| GS_IRS55[d] | 03:32:05.48 | -27:36:44.0 | $z \sim 1$ SF | 0 | 0.96 | 429.0 ± 7.3 | ... | ... | ... | 12.67 ± 2.27 | *4.70 ± 2.00* | *2.38 ± 2.00* |
| GS_IRS56 | 03:32:13.58 | -27:47:54.0 | $z \sim 2$ SF | 12 | 1.72 | 166.0 ± 5.9 | ... | 1.36 ± 0.19 | 3.51 ± 0.76 | 4.70 ± 2.26 | 5.54 ± 1.96 | 1.09 ± 4.48 |
| GS_IRS57 | 03:32:18.73 | -27:46:27.0 | $z \sim 2$ SF | 6 | 1.72 | 234.0 ± 8.3 | ... | 1.30 ± 0.27 | 2.84 ± 1.17 | 9.35 ± 2.20 | 8.23 ± 2.53 | 6.70 ± 2.00 |
| GS_IRS58[f] | 03:32:40.24 | -27:49:49.0 | $z \sim 2$ SF | 0 | 1.85 | 243.0 ± 8.7 | ... | 1.27 ± 0.14 | 3.51 ± 0.99 | 8.85 ± 2.28 | 11.15 ± 2.14 | 5.73 ± 2.88 |
| GS_IRS59[d] | 03:31:21.48 | -27:41:47.0 | $z \sim 1$ SF | 0 | 0.94 | 251.0 ± 8.9 | 2.57 ± 0.72 | ... | ... | 8.20 ± 2.37 | ... | ... |
| GS_IRS60 | 03:32:40.05 | -27:47:55.0 | $z \sim 2$ SF | 40 | 2.02 | 379.0 ± 6.6 | ... | 4.34 ± 0.27 | 9.79 ± 0.87 | 12.87 ± 2.54 | 12.39 ± 3.03 | 5.54 ± 3.12 |
| GS_IRS61 | 03:32:43.45 | -27:49:01.0 | $z \sim 2$ SF | 0 | 1.77 | 245.0 ± 8.7 | ... | 3.21 ± 0.16 | 5.56 ± 1.08 | 5.76 ± 2.20 | 3.43 ± 2.58 | 4.96 ± 2.80 |
| GS_IRS62 | 03:32:22.48 | -27:49:35.0 | $z \sim 1$ SF | 0 | 0.73 | 461.0 ± 7.8 | 3.31 ± 0.62 | 11.84 ± 0.22 | 29.30 ± 0.93 | 43.00 ± 2.52 | 37.17 ± 2.33 | 15.10 ± 5.64 |
| GS_IRS63 | 03:32:22.59 | -27:44:25.9 | $z \sim 1$ SF | 0 | 0.74 | 549.0 ± 7.5 | 1.64 ± 0.75 | 8.75 ± 0.30 | 18.06 ± 1.06 | 14.86 ± 2.62 | 16.56 ± 2.75 | 8.11 ± 3.22 |
| GS_IRS64[e] | 03:32:17.28 | -27:49:08.0 | $z \sim 2$ SF | 14 | 1.62 | 78.5 ± 5.2 | ... | 1.58 ± 0.39 | 5.28 ± 0.91 | 7.20 ± 2.28 | 8.19 ± 2.15 | 6.20 ± 2.00 |
| GS_IRS65[d] | 03:31:25.30 | -27:39:18.5 | $z \sim 1$ SF | 3 | 1.46 | 111.0 ± 6.7 | ... | ... | ... | 12.37 ± 2.39 | 10.87 ± 2.44 | 4.60 ± 2.00 |
| GS_IRS66[e] | 03:31:29.79 | -27:46:14.5 | $z \sim 1$ SF | 0 | 1.60 | 134.0 ± 4.5 | ... | ... | ... | 7.46 ± 2.27 | 10.30 ± 2.26 | 4.00 ± 2.00 |
| GS_IRS67[e] | 03:32:08.53 | -27:46:48.3 | $z \sim 1$ SF | 0 | 0.31 | 1050.0 ± 10.2 | 13.90 ± 0.68 | 31.66 ± 0.62 | 49.78 ± 1.48 | 29.98 ± 2.04 | 14.87 ± 2.70 | 8.02 ± 6.62 |
| GS_IRS68[e] | 03:32:13.24 | -27:42:40.9 | $z \sim 1$ SF | 0 | 0.60 | 706.0 ± 7.6 | 10.30 ± 0.69 | ... | ... | 13.90 ± 3.32 | 7.61 ± 4.38 | *3.10 ± 2.00* |
| GS_IRS69[e] | 03:32:24.84 | -27:47:06.5 | $z \sim 1$ SF | 0 | 0.36 | 207.0 ± 7.4 | 2.91 ± 0.54 | 5.26 ± 0.20 | 5.46 ± 1.20 | 5.76 ± 2.40 | 3.75 ± 2.65 | ... |
| GS_IRS70[e] | 03:32:27.71 | -27:50:40.6 | $z \sim 1$ SF | 25 | 1.10 | 599.0 ± 7.9 | 2.70 ± 0.50 | 7.91 ± 0.34 | 13.39 ± 0.72 | 17.05 ± 2.50 | 12.94 ± 2.95 | 5.60 ± 5.18 |
| GS_IRS71[e] | 03:32:32.99 | -27:50:30.1 | $z \sim 1$ SF | 0 | 0.67 | 370.0 ± 6.4 | 3.29 ± 0.61 | 7.38 ± 0.23 | 13.77 ± 1.28 | 16.14 ± 2.50 | 10.43 ± 2.93 | 5.76 ± 2.66 |
| GS_IRS72[e] | 03:32:38.75 | -27:47:32.2 | $z \sim 1$ SF | 2 | 0.46 | 782.0 ± 8.4 | 6.33 ± 0.57 | 11.38 ± 0.23 | 8.65 ± 0.69 | 6.14 ± 3.64 | 3.61 ± 3.05 | *0.70 ± 2.00* |
| GS_IRS73 | 03:32:43.24 | -27:47:56.2 | $z \sim 1$ SF | 0 | 0.67 | 162.0 ± 5.7 | ... | 4.20 ± 0.33 | 10.05 ± 1.03 | 14.74 ± 2.76 | 10.98 ± 3.08 | 4.19 ± 2.78 |
| GS_IRS74 | 03:32:44.32 | -27:49:11.9 | $z \sim 1$ SF | 0 | 2.00 | 268.0 ± 5.8 | ... | 1.40 ± 0.20 | 3.07 ± 0.90 | 9.41 ± 2.12 | 10.70 ± 2.40 | 4.00 ± 2.00 |
| GS_IRS75[e] | 03:32:47.30 | -27:42:47.9 | $z \sim 1$ SF | 43 | 0.74 | 471.0 ± 8.0 | 14.90 ± 0.79 | 24.48 ± 2.67 | 29.91 ± 5.78 | 23.56 ± 4.66 | 12.15 ± 5.20 | 7.66 ± 8.44 |
| GS_IRS76[e] | 03:32:48.83 | -27:42:35.0 | $z \sim 2$ SF | 33 | 1.96 | 226.0 ± 8.1 | ... | ... | ... | *12.30 ± 2.00* | *25.70 ± 2.00* | *23.20 ± 2.00* |
| GS_IRS77[d] | 03:32:49.33 | -27:54:27.2 | $z \sim 1$ SF | 16 | 0.56 | 119.0 ± 7.2 | ... | ... | ... | 5.89 ± 1.93 | 6.86 ± 2.00 | 2.51 ± 2.88 |
| GS_IRS78[e] | 03:32:54.34 | -27:41:30.3 | $z \sim 2$ SF | 0 | 2.00 | 358.0 ± 6.2 | ... | ... | ... | 13.60 ± 2.19 | 10.96 ± 2.34 | *2.00 ± 2.00* |
| GS_IRS79[d] | 03:33:14.96 | -27:45:56.6 | $z \sim 1$ SF | 9 | 0.53 | 118.0 ± 7.1 | 2.03 ± 0.83 | ... | ... | *3.00 ± 2.00* | *5.50 ± 2.00* | *0.70 ± 2.00* |
| GS_IRS80[d] | 03:32:36.52 | -27:46:30.7 | $z \sim 1$ SF | 0 | 1.05 | 267.0 ± 5.8 | 2.11 ± 0.51 | 5.40 ± 0.22 | 11.89 ± 1.33 | 17.14 ± 3.22 | 11.94 ± 2.43 | 8.31 ± 2.46 |
| GS_IRS81 | 03:32:38.49 | -27:46:31.9 | $z \sim 2$ SF | 38 | 2.55 | 183.0 ± 12.4 | ... | 4.48 ± 0.60 | 14.81 ± 1.44 | 23.34 ± 2.94 | 19.52 ± 2.30 | 10.98 ± 2.58 |

[a] When we perform spectral decomposition on each IRS spectrum, we calculate the percentage of mid-IR luminosity due to a power-law component (see §2.3). We attribute the power-law component to an AGN, so this percentage represents the strength of the AGN in the mid-IR luminosity.
[b] We calculate redshifts for each source using the IRS spectrum (see §2.4).
[c] For sources that do not have measurements at SPIRE wavelengths from the GOODS-Herschel prior catalog, we make a measurement from the SPIRE images directly even if it is not a formal detection so it can be included in the SED fitting (see §3). These additional measurements are identified by italic font. Entries which are blank indicate that the source was too blended in SPIRE to extract a flux measurement; this mostly affects 500 μm. We list only the instrument noise for each detection or measurement, but we add in the confusion noise when performing the far-IR SED fitting.
[d] These sources were rejected from the creation of the composites due to blending in the far-IR (see §3).
[e] Sources rejected from creation of the composites because they lack data in the wavelength range used to normalize the mid-IR spectra (see §3).
[f] Sources rejected from creation of the composites due to having a noisy spectrum (see §3).


\clearpage

\section{B. Correlations between modified blackbody parameters}
In the plots below, we show the correlation between parameters when fitting our two-temperature modified blackbody model from Equation (1).
We show the results from our Monte Carlo simulations (see \S \ref{sec:sed}) for the $z\sim2$ SF galaxy subsample as an illustrative case; the results are similar for the other three subsamples. We overplot the mean
value of each parameter as the red cross. The plots below show that certain parameters are highly correlated (e.g., the temperatures of the warm and cold components) while others are not (e.g., the cold temperature and normalization). The errors we report for the dust temperatures and IR luminosities include the covariance between the parameters but we note that the error is dominated by the diagonal elements of the covariance matrix. The warm and cold temperatures, though correlated, do not overlap in their respective range of values, giving us confidence that we are clearly measuring two separate temperature components. 
\begin{figure}[ht!]
\plottwo{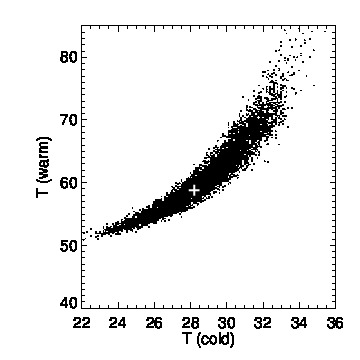}{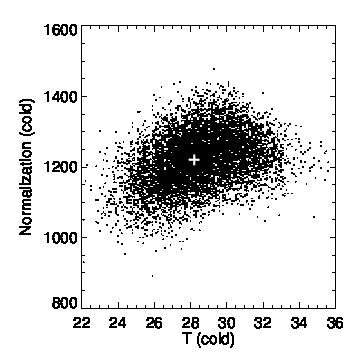}
\end{figure}
\begin{figure}
\plottwo{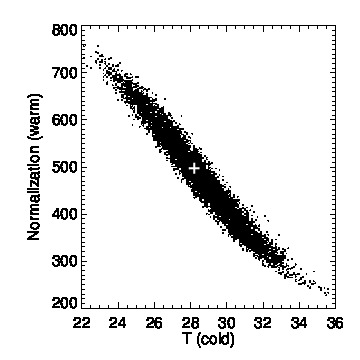}{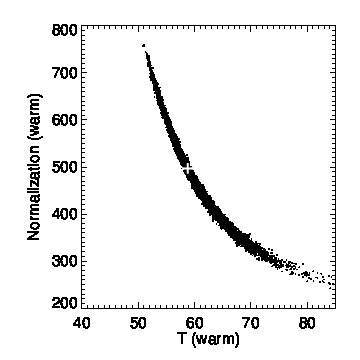}
\end{figure}
\begin{figure}
\plottwo{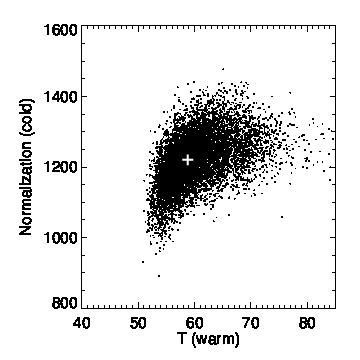}{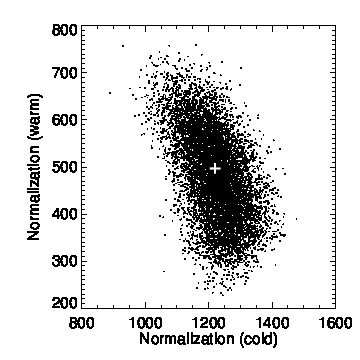}
\end{figure}

\end{document}